\documentclass[superscriptaddress, amsmath,amssymb, showpacs, pre, twocolumn]{revtex4-1}
\usepackage{graphicx}
\usepackage{dcolumn}
\usepackage{bm}
\usepackage{hyperref}
\usepackage{color}
\usepackage{amsmath}
\usepackage{amssymb}
\usepackage[caption=false]{subfig}
\begin{document}

\title{Dynamically Generated Patterns in Dense Suspensions of Active Filaments}
\author{K.~R.~Prathyusha}
\affiliation{School of Science and Engineering, University of Dundee, Dundee, DD1 4HN, United Kingdom}%
\author{Silke Henkes}
\affiliation{Institute of Complex Systems and Mathematical Biology, Department of Physics, University of Aberdeen, Aberdeen AB24 3UE, United Kingdom}
\author{Rastko Sknepnek}%
\affiliation{School of Science and Engineering, University of Dundee, Dundee, DD1 4HN, United Kingdom}%
\affiliation{School of Life Sciences, University of Dundee, Dundee, DD1 5EH, United Kingdom}
\email{r.sknepnek@dundee.ac.uk}

\date{\today}

\begin{abstract}
We use Langevin dynamics simulations to study dynamical behaviour of a dense planar layer of active semi-flexible filaments. Using the strength of active force and the thermal 
persistence length as parameters, we map a detailed phase diagram and identify several non-equilibrium phases in this system. In addition to a slowly flowing melt phase, we observe that 
for sufficiently high activity, collective flow accompanied by signatures of local polar and nematic order appears in the system. This state is also characterised by strong density fluctuations. 
Furthermore, we identify an activity-driven cross-over from this state of coherently flowing bundles of filaments to a phase with no global flow, formed by individual filaments coiled into rotating 
spirals. This suggests a mechanism where the system responds to activity by changing the shape of active agents, an effect with no analogue in systems of active particles without internal 
degrees of freedom.  
\end{abstract}


\maketitle

\section{Introduction}
\label{sec:introduction}

Processes that involve active motion and remodelling of protein filaments are particularly important for the functioning of a cell \cite{alberts-thecell-08}. In particular, the dynamics of cytoskeletal actin filaments 
are characterised by a constant supply of mechanical energy by myosin motor proteins, which hydrolyse ATP to slide along actin filaments~\cite{fletcher2010cell,huber2013emergent}. 
This directed motion is integral to cell migration~\cite{warrick1987myosin,ridley2003cell}. 
Motility assays~\cite{sheetz1983movement,sheetz1984atp} have served as simple yet elegant model systems for \emph{in-vitro} studies of many active cellular processes. In motility assays, 
filaments are driven by molecular motors that are typically grafted to a flat substrate, with the energy required for their active motion supplied by ATP. The basic design of motility assays enables 
a reasonably accurate control of the key parameters, which would be very hard to achieve in \emph{in-vivo}. Despite their simplicity compared 
to actual cells, actively driven \emph{in-vitro} filaments exhibit fascinating self-organised motion patterns \cite{ndlec1997self,surrey2001physical,e2011active,koster2016actomyosin,linsmeier2016disordered}.
Understanding and characterising these patterns has been a focus of active research~\cite{kruse2004asters,sriram-10,marchetti-13,murrell2015forcing}.


Only several studies to date have investigated the collective motion and pattern formation in the high density regime. For example, Sumino et al.~\cite{sumino2012large} reported the existence of an 
intriguing vortex state in a motility assay of microtubules driven by dynein motor proteins. Large  vortices with an average diameter exceeding the mean length of individual filaments by more than an
order of magnitude form a lattice structure on sufficiently long time scales. Furthermore, working with actin filaments propelled by heavy meromyosin motors attached to a coverslip, 
Schaller et al.~\cite{schaller2010polar} identified several interesting collective motion patterns as a function of the actin density.

Interesting collective behaviour of actively driven filamentous systems is not restricted to motility assays~\cite{winkler2017active}. Equally fascinating motion patterns arise in low-density mixtures of microtubules suspended 
at the oil-water interface and propelled by ATP driven kinesin motors~\cite{tsanchez-12, keber2014topology, decamp2015orientational}, where flow is accompanied by spontaneous 
generation~\cite{tsanchez-12} of motile topological defects (so-called active turbulence \cite{giomi2013defect,thampi2013velocity}) and the development of orientational order of those
defects~\cite{decamp2015orientational}. Many of the qualitative features observed in these experiments have been reproduced using continuum models for active 
nematics~\cite{giomi2013defect,giomi2014defect,thampi2013velocity,thampi2014instabilities}. Finally, in an experiment by Keber et al.~\cite{keber2014topology}, the microtubule/kinesin
mixture was suspended onto the surface of a nearly spherical lipid vesicle. The presence of non-zero curvature leads to an even richer set of collective motion patterns that are only 
partly understood~\cite{alaimo2017curvature,henkes2017dynamical}. 

Finally, bacterial colonies provide another example of elongated active agents with rich collective behaviour. 
Myxobacteria, for example, form a striking variety of collective motion patterns~\cite{kaiser2003coupling} without any long-range chemical signals, but solely due to short-range steric effect 
and rod-like cell shape~\cite{peruani2012collective}.

Many of the non-equilibrium patterns observed in motility assays, such as asters and vortices can be described by the active gel theory~\cite{kruse2004asters,juelicher2007active,joanny2009active,prost2015active}. 
Continuum description of the active gel theory have been augmented by studies of a number of numerical models with different levels of microscopic details~\cite{gibbons2001dynamical,nedelec2002computer,kraikivski2006enhanced,kim2009computational,jung2015f,popov2016medyan}. 
Many of those models are, however, rather simplified and assume stiff, rod-like filaments and/or ignore steric effects. In addition, the effects of hydrodynamics interactions mediated by the flow of the fluid surrounding 
filaments are also often omitted.

In the dilute regime, steric interactions play a limited role and insights can be gained by studying individual filaments~\cite{gayathri-12,hiang-14a,riseleholder-15,laskar2015brownian,kaiser2015does} 
or hydrodynamic equations derived from microscopic models~\cite{baskaran2008enhanced,peshkov2012nonlinear,gao2015multiscale}. Studies of individual active filaments either 
pivoting or freely swimming, showed that activity can drive conformational transformations~\cite{sekimoto-95}, such as spiralling and spontaneous 
beating~\cite{bourdieu-95,rchelakkot-13,riseleholder-15}, both in the presence and absence of hydrodynamic interactions. Balancing the bending moment with the torque 
produced by the active force shows~\cite{bourdieu-95,rchelakkot-13} that activity increases the tendency 
of the filament to buckle, thus reducing its persistence length. Similar effects have also been observed in models that treat activity as time-correlated random
 forces~\cite{tbliverpool-03,aghosh-14,eisenstecken2016conformational}.  
 
It is important to note that most microscopic derivations of continuum equations assume the dilute limit and only consider binary collisions between filaments. For dense systems, 
considering only binary collisions is not justifiable and it is no longer possible to relate parameters of the microscopic model to the parameters in the effective continuum theory. 
Numerical simulations are, therefore, crucial to understand the collective behaviour of dense active systems. 

In this paper, we explore the collective motion patterns of a dense planar layer of active semi-flexible filaments. We use Langevin dynamics simulations to study out of equilibrium 
behaviour of a bead-spring filament model in the presence of steric repulsion. Inspired by the activity mechanism of motility assays, the driving force acts along the contour of the filament. 
Our model includes frictional coupling to the environment, but ignores the effects of flow. While the importance of hydrodynamic effects in motility assay experiments it still under debate, 
 the experiments of Sumino et al.~\cite{sumino2012large} suggest that omitting long-range hydrodynamic interactions in these systems is justifiable. This also significantly reduces the high computational cost associated with simulating significant system sizes with this model. 

We map a non-equilibrium phase diagram as a function of activity and stiffness of filaments and identify five distinct phases: A polymer \emph{melt} phase is followed by a 
\emph{flowing melt} region with signatures of both polar and nematic symmetries as activity increases. For stiffer filaments, further increase in activity leads to a phase 
\emph{segregated} state akin to the motility induced phase segregation observed for isotropic active particles~\cite{cates2015motility} (MIPS). For flexible filaments, 
increase in activity leads to a \emph{swirling} state. Finally, for high activities we observe a \emph{rotating spiral} phase where filaments adjust their conformation to accommodate the activity. 
This is contrary to the behaviour observed in models of structureless active particles where (MIPS) becomes more prominent with 
increasing activity~\cite{redner2013structure}. 

The paper is organised as follows. In Sec.~\ref{sec:model} we discuss a coarse-grained model for semi-flexible filaments subject to an active force acting along the contour of each
filament. In Sec.~\ref{sec:results} we present and discuss results of detailed Langevin dynamics simulations and map the non-equilibrium phase diagram. Finally, in Sec.~\ref{sec:summary}
we summarise our main findings, comment on potential experimental realisations and discuss how the model could be extended to describe specific experiments. In Appendix~\ref{app:langevin_vs_brownian}
we compare Langevin dynamics simulations with the more commonly used Brownian dynamics studies.

\section{Model}
\label{sec:model}
\begin{figure}[tb]
\centering
\includegraphics[width=0.75\columnwidth]{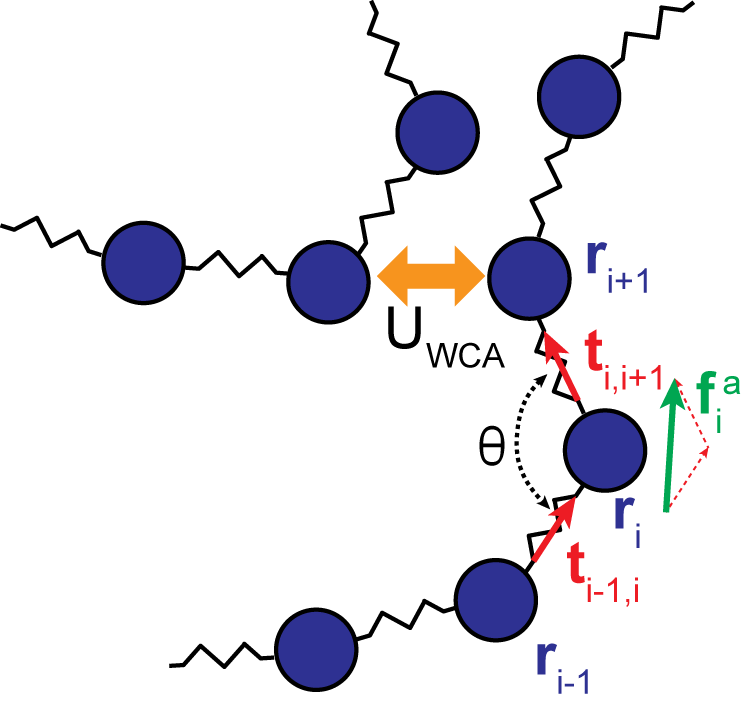}
\caption{(Colour online) Schematic representation of the model for self-propelled semi-flexible filaments. Spherical beads of diameter $\sigma$ are connected through 
harmonic springs of spring constant $k_b$. The filament flexibility is modelled as a bending penalty for angle $\theta$ with stiffness constant $\kappa$. Beads
belonging to different chains, as well as beads on the same chain that are more than two beads apart, interact via a short-range, repulsive interaction, modelled with the Weeks-Chandler-Andersen potential.
Finally, each bead is subject to an active force $\mathbf{f}_{i}^{a}$ pointing along the tangent to the filament contour at $i$. Note that for clarity, distances between beads along the filament
have been exaggerated. \label{fig:model}}
\end{figure}

We adapt the model recently used by Isele-Holder et al.~\cite{riseleholder-15} to study conformations of a single semi-flexible filament under the influence of an active
force of constant magnitude acting along its contour (FIG.~\ref{fig:model}). Our system consists of $M$ such filaments confined to a periodic square region of length $L_{\text{box}}$. 
Each filament is modelled as a chain of $N$ beads of diameter $\sigma$, so that all filaments have the same degree of polymerisation. In most simulations, the packing fraction 
$\phi = (MN\sigma^2\pi)/4 L_{\text{box}}^2$ was set to $\phi\approx0.65$. This particular choice 
of the packing fraction ensures that the system is dense on one hand, but on the other hand, it remains below both the packing fraction of the triangular lattice and random close packing
In other words, there is nothing special about $\phi\approx0.65$, except that it is high enough that steric effects cannot be ignored but also low enough for the system to be 
sufficiently far away from being jammed.

\begin{figure*}[t] 
\centering
\includegraphics[width=\textwidth]{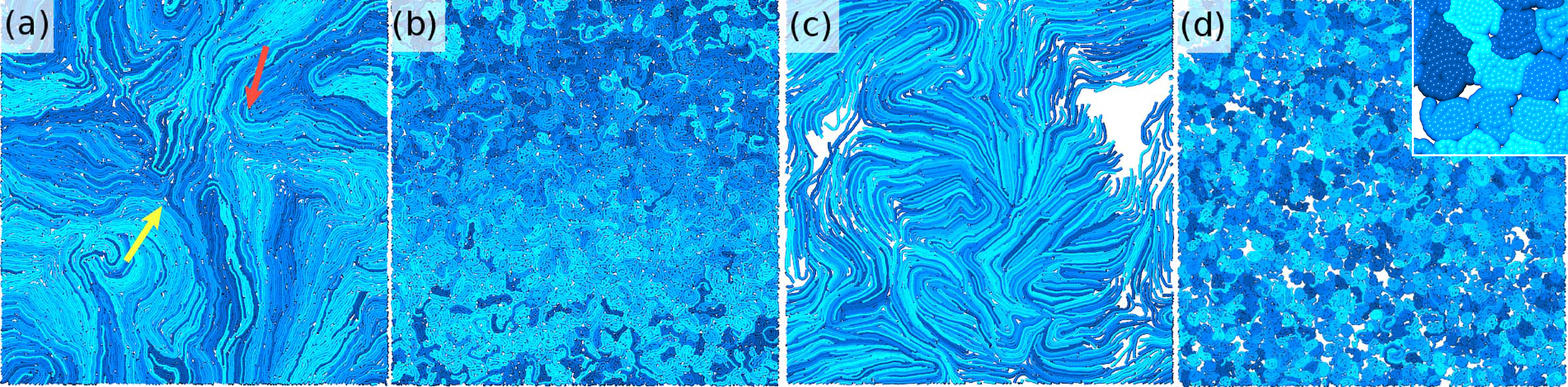} 
\caption{(Colour online) Snapshots of simulations for $N=50$ in four non-equilibrium phases: (a) flowing melt ($\xi_p/L=0.2$, $Pe=35.5$), (b) swirl ($\xi_p/L=0.04$, $Pe=177.58$), 
(c) segregated phase ($\xi_p/L=0.82 $, $Pe=355.16$) and (d) spirals ($\xi_p/L=0.04$, $Pe=17757.8$). Individual filaments are marked by a different shade. Arrows in (a) point to examples 
of $+1/2$ (red) and $-1/2$ (yellow) topological defects. White areas in (c) are regions devoid of filaments, indicating substantial density fluctuations. Inset in panel (d) is a zoom-in on several spirals. 
\label{fig:phase} }
\end{figure*}

Interactions between beads are modelled with bonded and short-range non-bonded pair potentials, i.e., $U=U_{B}+U_{NB}$. Bonded interactions, $U_B=U_s+U_b$, account for both chain stretching, 
modelled with the FENE bond potential \cite{kkremer-90}, 
\begin{equation}
U_{s}\left(r_{ij}\right)=-1/2k_bR_0^2\ln\left(1-\left({r_{ij}}/{R_0}\right)^2 \right)
\end{equation}
and bending, modelled with the standard harmonic angle potential \cite{rapaport2004art},
\begin{equation}
U_{b}=\kappa\left(\theta -\pi \right)^2.
\end{equation}
Here, $\kappa$ is the bending stiffness, $\theta$ is the angle between three consecutive beads, $R_0=1.3\sigma$ is the maximum bond length, $k_b=3300k_BT/\sigma^2$ 
is the bond stiffness (making the chain effectively non-stretchable), and $r_{ij}\equiv\left|\mathbf{r}_{ij}\right|=\left|\mathbf{r}_i-\mathbf{r}_j\right|$ is the distance between beads
at positions $\mathbf{r}_i$ and $\mathbf{r}_j$. For a stiff filament, the bending stiffness $\kappa$ in the discrete model is related to the continuum bending stiffness, $\tilde{\kappa}$, as
$\kappa \approx \tilde{\kappa}/2b$, where $b$ is the average bond length. Non-bonded interactions, $U_{NB}$, account for steric repulsions and are modelled with the 
Weeks-Chandler-Anderson potential \cite{weeks1971role}, 
\begin{equation}
U_{WCA}(r_{ij})=4\varepsilon\left[\left(\frac{\sigma}{r_{ij}}\right)^{12}-\left(\frac{\sigma}{r_{ij}}\right)^{6}+\frac{1}{4}\right],
\end{equation}
where $\varepsilon$ measures the strength of steric repulsion. These parameters lead to $b\approx0.86\sigma$, which ensures that filaments do not intersect. Finally, $L=(N-1)b$
is the mean filament length and $T$ is set to $0.1\varepsilon/k_B$ in all simulations. 

The active force on bead $i$ is modelled as  
\begin{equation}
\mathbf{f}^a_i=f_p(\mathbf{t}_{i-1,i}+\mathbf{t}_{i,i+1}),
\label{eq:active_force}
\end{equation}
which mimics active driving produced by a homogeneous distribution of molecular motors on the substrate underneath the filaments. Here, $f_p$ is the strength of the force and 
$\mathbf{t}_{i,i+1}=\mathbf{r}_{i,i+1}/r_{i,i+1}$ is the unit-length tangent vector along the bond connecting beads $i$ and $i+1$; end beads have only contributions from one neighbouring bond. 
We note that the active force in Eq.~(\ref{eq:active_force}) is slightly different than the active force used in Ref.~\cite{riseleholder-15}, where the tangent vectors $\mathbf{t}_{ij}$ were 
not normalised. Given that filaments are effectively unstretchable, the difference between the two expressions for active force is just a scaling factor of order one.
We point out that an important, but unsolved question is how important the precise microscopic driving mechanism is for long-range collective behaviour. Here we simply note that the extensile 
nematic system, such as in the experiments of Sancez et al.~\cite{tsanchez-12}, is driven by \emph{pair} forcing of filaments opposite to each other~\cite{gao2015multiscale}. 
\emph{Unidirectional} forcing, as implemented here, has been shown by two of us to lead to a mix of polar and nematic properties \cite{henkes2017dynamical}, a result that we recover here.

In experiments, filaments are surrounded by a fluid that mediates long-range hydrodynamic interactions. Here we consider the \emph{dry} limit, where the damping from the medium 
dominates over hydrodynamic interactions and the surrounding fluid only provides single-particle friction. In this regime, the filament dynamics is described by Langevin equations of motion:
\begin{equation}
m_i\ddot{\mathbf{r}}_i = \mathbf{f}^a_i -\gamma\dot{\mathbf{r}}_i - \nabla_{\mathbf{r}_i}U(r_{ij}) + \mathbf{R}_i(t),\label{eq:Langevin}
\end{equation}
where $m_i=1$ is the mass of bead $i$. Time is measured in units of $\tau=\sqrt{m\sigma^2/\varepsilon}$. $\mathbf{R}_i(t)$ is a delta-correlated random force with zero mean and, as
required by the fluctuation-dissipation theorem, variance $\langle \mathbf{R}_{i}\left(t\right)\cdot\mathbf{R}_{j}\left(t^\prime\right)\rangle=\langle R^{x}_{i}\left(t\right)R^{x}_{j}\left(t^\prime\right)\rangle+\langle R^{y}_{i}\left(t\right)R^{y}_{j}\left(t^\prime\right)\rangle=4\gamma k_BT \delta_{ij}\delta(t-t')$, where $\gamma$ is the friction coefficient, and the prefactor $4$ reflects the fact that the system is
confined to two dimensions.  Eqs.~(\ref{eq:Langevin}) were integrated with time step $\Delta t=10^{-3}\tau$ using 
LAMMPS~\cite{splimpton-95}, with an in-house modification to include the active propulsion force. A typical configuration contained $\approx5\times10^4$ beads and was generated by 
placing chains at random into the simulation box ($L_{\text{box}}=250\sigma$), making sure that there were no intersections. This configuration was then relaxed for $10^4\tau$, followed by a 
production run of $10^5\tau$. 

We briefly comment on the use of Langevin dynamics. Mesoscopic active systems are typically at very low Reynolds numbers and inertial effects are negligible. Therefore, 
most models from the outset assume the overdamped limit and use first order equations of motion, where the mass term is omitted. However, there is no a priori reason that prohibits the 
inertial term from being retained. In dense systems with steep short-range repulsive interactions, retaining it turns out to be computationally beneficial as it allows the use of a significantly larger 
simulation time step~\cite{leimkuhler2015molecular} compared to what is possible with the Brownian dynamics approach. We exploit the fact that there is a separation of time scales between individual 
bead collisions and the mesoscopic dynamics. As we show in Appendix~\ref{app:langevin_vs_brownian}, the behaviour of the system at the time scales associated with collective flow is identical
regardless of whether Langevin or Brownian dynamics is used. 

Our key parameters are the degree of polymerisation, $N\in\{5, 10, 25, 50\}$, the filament stiffness, $\kappa=1-20k_BT$ and the magnitude of the active force, $f_p=2\times10^{-4}-1\:k_BT/\sigma$. 
We identify two dimensionless numbers: the relative filament stiffness, or scaled persistence length,
\begin{equation}
\frac{\xi_p}{L} = \frac{\tilde{\kappa}}{Lk_BT}=\frac{2b\kappa}{Lk_BT}
\end{equation}
where $\xi_p=2b\kappa/k_BT$ is the thermal (i.e., passive) persistence length, and the active P{\'e}clet number,
\begin{equation}
 Pe=\frac{f_pL^2}{\sigma k_BT}.
 \end{equation}
 As previously shown~\cite{sekimoto-95,rchelakkot-13}, the flexure number, $\mathfrak{F}=f_pL^3/\kappa=\sigma^2 Pe/(\xi_p/L)$, controls the buckling instabilities of a single active filament. 
 Meanwhile, the P{\'e}clet number controls the onset of the MIPS state in systems of isotropic active brownian particles~\cite{redner2013structure}.

\begin{figure}[t]
\centering
\includegraphics[width=0.95\columnwidth]{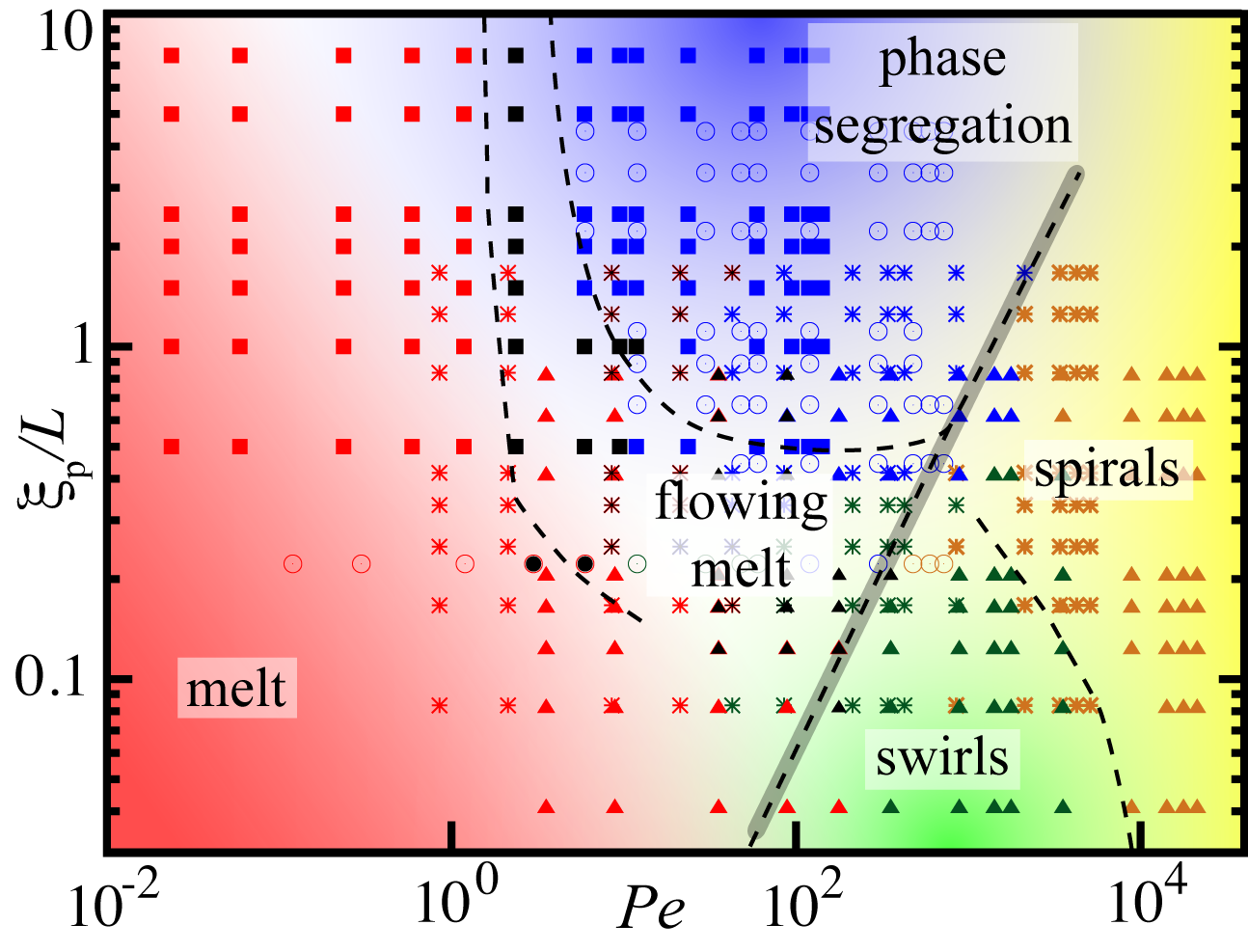}
\caption{(Colour online) Non-equilibrium phase diagram for $\xi_p/L$ vs.~$Pe$ at packing fraction $\phi = 0.64905$. Results for different filament lengths are plotted on the same graph
since they show the same scaling. Symbols mark individual simulations for different filament lengths: $N=5$ ($\blacksquare$), $N=10$ ($\bigcirc$), $N=25$ ($\ast$) and $N=50$ 
($\blacktriangle$), with phases coded by colour. Different phases were identified by visually inspecting each individual run. Crossovers between phases are indicated by shading. 
Dashed lines indicate rough boundaries between phases and serve as a guide to the eye. The grey-shaded bar indicates the threshold to spiralling, set by the flexure 
number $\mathfrak{F}\approx 10^3$  \cite{riseleholder-15}.
\label{fig:phase-diagram}}
\end{figure}

\section{Results and Discussion}
\label{sec:results}
We start by constructing a detailed non-equilibrium phase diagram for this system and proceed to describe and characterise several different motion patterns. While there are similarities
with the behaviour of a single active filament \cite{riseleholder-15}, the combination of self-avoidance and flexibility in dense suspensions leads to interesting collective behaviour even in 
the absence of any explicit aligning interactions. In FIG.~\ref{fig:phase-diagram} we show the phase diagram in the $\xi_p/L$ vs.~$Pe$ plane with snapshots of the simulations in the four 
main non-equilibrium phases shown in FIG.~\ref{fig:phase} (see also movies in the Supplemental Materials \cite{supporting}). It is also important to note that the phase behaviour of the system 
can be fully described by the two dimensionless numbers, $Pe$ and $\xi_p/L$.

\begin{figure}[tb] 
\centering
\includegraphics[width=\columnwidth]{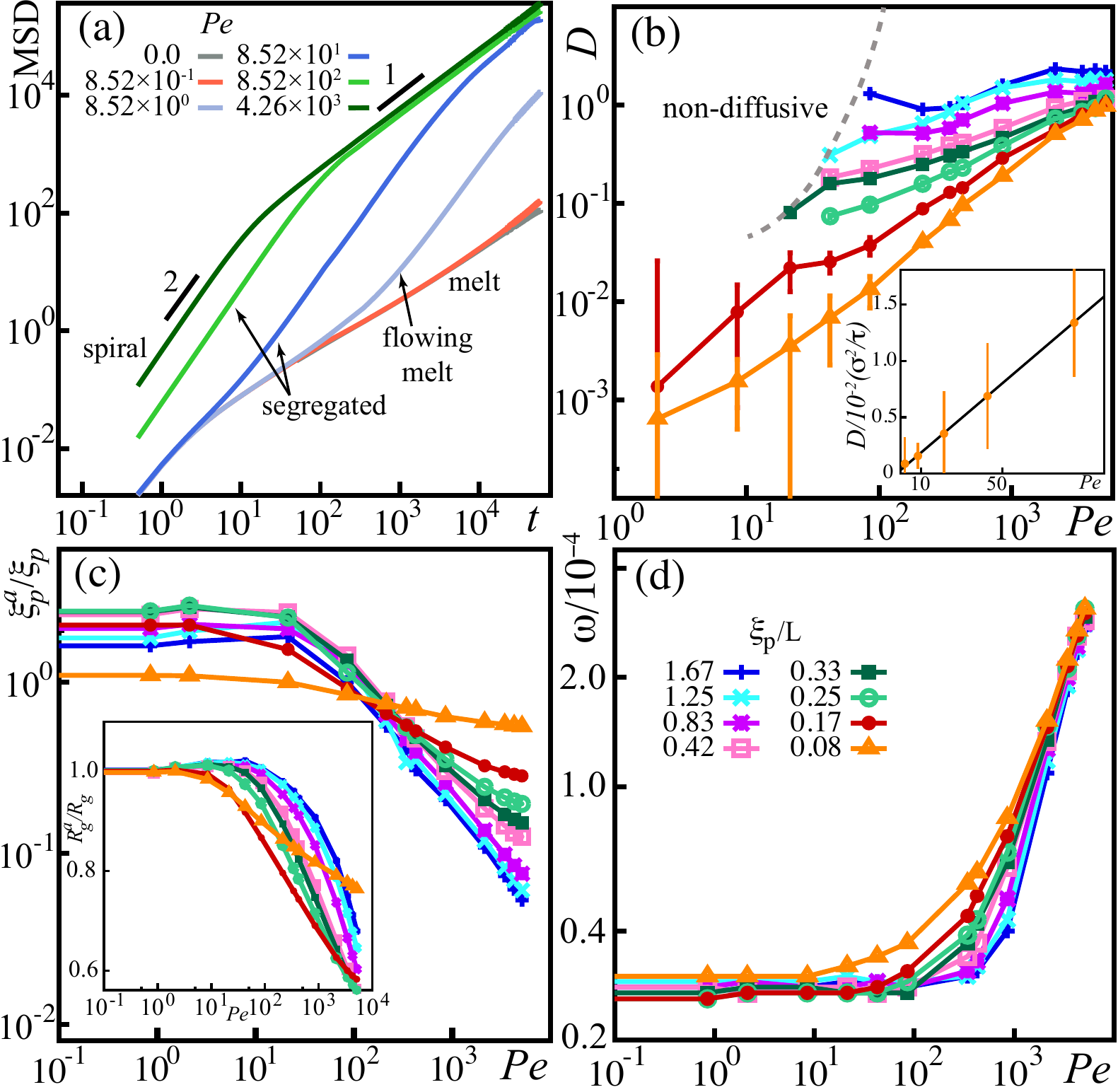} 
\caption{(Colour online) (a) Mean-square displacement (MSD) as a function of time for $\xi_p/L\approx0.83$ for six values of $Pe=0, 0.852, 8.52, 85.2, 852, 4260$. (b) Log-log plot of the diffusion 
coefficient as a function of $Pe$ in the regime where the system is diffusive. \emph{Inset:} Linear scaling at low activity for a soft ($\xi_p/L=0.08$) filament. (c) Ratio of active ($\xi_p^a$) to passive
($\xi_p=2\kappa b/k_BT$) persistence length for a range of values of bare bending stiffnesses. Note that due to steric effects  the measured persistence length at low activity is always larger 
than $\xi_p$. \emph{Inset:} Ratio of active, $R_g^a$, to passive, $R_g$, radius of gyration as a function of $Pe$.  (d) Average angular velocity of rotation $\omega$ defined in Eq. \ref{eq:omega}
around centres of mass of filaments as a function of $Pe$. Legend in (d) also applies to panels (b) and (c). $N=25$ in all plots. }
\label{fig:struct-dynamic}
\end{figure}

In order to characterise the nature of the flow in the system we computed the mean squared displacement (MSD) of the centre of mass of filaments, averaged over all filaments 
(FIG.~\ref{fig:struct-dynamic}a).  Without activity (i.e., for $Pe=0$), after the usual short time ballistic relaxation, the MSD curve shows subtly sub-diffusive dynamics with 
exponent $\approx 0.8$ 
indicating slow dynamics resembling supercooled liquids. As expected, activity introduces flow, leading to the \emph{melt} phase. For low activity, steric effects dominate 
and the system resembles a conventional polymer melt where activity acts only as a weak perturbation. It is interesting to note that for flexible filaments 
($\xi_p/L\lesssim0.1$) at $Pe\gtrsim1$ the diffusion coefficient increases linearly with P{\'e}clet number 
(inset, FIG. \ref{fig:struct-dynamic}b). While it is not surprising that the diffusion coefficient increases with activity (see also Ref.~\cite{eisenstecken2016conformational}), the origin of the linear dependence 
on the $Pe$ is at present not clear.

Stiffer filaments, on the other hand, exhibit a transition from subdiffusive to superdiffuisive flow at $Pe\approx1$. For longer filaments, in this \emph{flowing melt} regime, one observes $\pm1/2$ topological
defects (FIG.~\ref{fig:phase}a) reminiscent of those seen in the experiments of Sanchez et al.~\cite{tsanchez-12}. However, it is important to note that there are two important differences between 
the structure of the defects observed in our simulations and those seen in the experiments of Sanchez et al. First, experiments are performed at much lower densities than studied here and,
 second, in the experiments individual microtubules only slightly bend, nowhere near the fully bent hair-pin configuration observed in this study. It is also still not completely clear if long-ranged hydrodynamics
interactions have an important effect on the dynamics in the experimental system. 

In our simulations, defect motion is very slow and is driven by filaments sliding along their contour. It is straightforward to properly identify and track the defects, e.g. by using a version of the algorithm 
proposed by Zapotocky et al.~\cite{zapotocky1995kinetics,henkes2017dynamical}, as outlined in Appendix \ref{app:defects}.
However, at the time scales accessible to our simulations, defects move over very short distances, insufficient to extract meaningful information about their dynamics. Therefore, we did not attempt
to study the defect dynamics in detail, but instead we applied the defect finding algorithm only to several selected snapshots of the simulation to showcase their existence. We note that defect motion 
in self-propelled systems appears to generally be much slower than individual particle motion~\cite{henkes2017dynamical, shi2013topological}.

\begin{figure}[tb] 
\centering
\includegraphics[width=0.85\columnwidth]{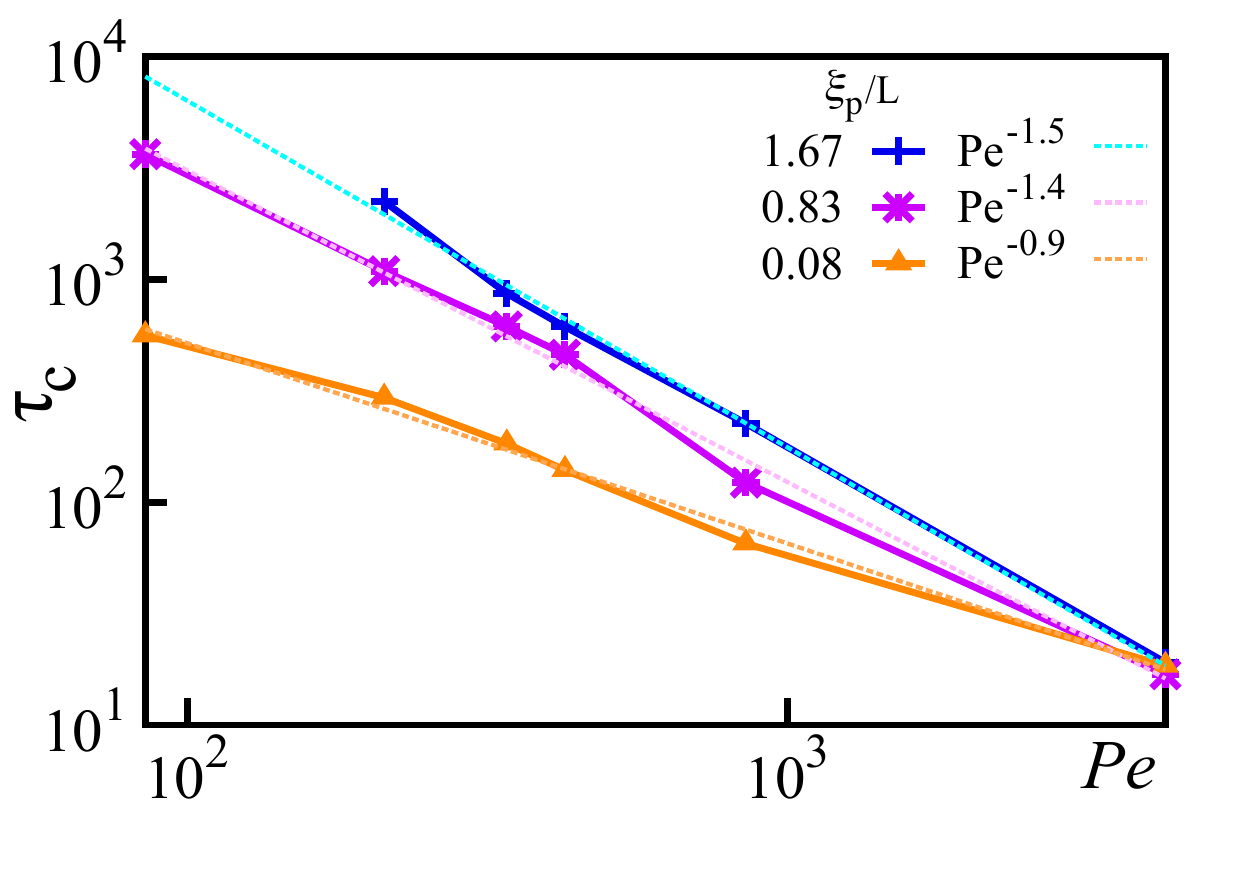} 
\caption{(Colour online) Transition time from super diffusive to diffusive behaviour in the segregated phase as a function of P{\'e}clet number for $\xi_p/L=0.08,0.83$ and $1.67$. }
\label{fig:tc_fall}
\end{figure}

As the filament stiffness increases ($\xi_p/L\gtrsim0.2$), large bends become costly and at sufficiently high P{\'e}clet numbers, $Pe \gtrsim10$, similar to the P{\'e}clet threshold in MIPS of self-propelled 
particles~\cite{redner2013structure}, the system crosses over into a \emph{phase-segregated} state (FIG.~\ref{fig:phase}c) characterised by filaments aligning and flowing as a coherent bundle. The mechanism
 that leads to this behaviour can be explained as follows. When multiple filaments collide with each other in a head-to-front manner, steric interactions align them into bundles that propagate 
 coherently~\cite{hiang-14a}. This, however, happens only if filaments are sufficiently stiff since for soft filaments, fluctuations in the direction of motion prevent them from bundling~\cite{supporting}. Bundles 
 break up due to collisions with filaments in surrounding bundles and thermal noise. Typically, we observe several large clusters moving in different directions. The cluster size depends on the density
 of filaments (in the dilute regime this aggregation is absent; see also movies in~\cite{supporting}). The motion of the bundles is accompanied by substantial density fluctuations (white regions in FIG.~\ref{fig:phase}c), 
akin to those seen in the MIPS phase of self-propelled disks and rods. Once formed, the entire bundle retains its direction over an extended period of time, in some cases comparable to the length of the entire 
simulation. This can be seen in the MSD curves (FIG.~\ref{fig:struct-dynamic}a), which show wide regions of persistent, $\sim t^2$, behaviour. At long times, however, the direction of motion of bundles 
decoheres and diffusive behaviour is recovered. It is interesting that the transition time, $\tau_c$, of the onset of diffusive behaviour \emph{reduces} with activity as $\tau_c\sim (Pe)^{-\zeta}$, with 
$\zeta\approx0.9-1.5$, where the exponent $\zeta$ increases with filament stiffness (FIG.~\ref{fig:tc_fall}). A similar activity-dependent decrease of $\tau_c$ has been observed in single-filament 
studies~\cite{riseleholder-15}. 

The \emph{phase-segregated} state is not stable for large P{\'e}clet numbers. 
For $Pe\approx10^2-10^4$ and $\xi_p/L\lesssim0.2$ the system enters the \emph{swirling} state (FIG.~\ref{fig:phase}b), in which the dynamics is dominated by large fluctuations of filament shapes with 
individual filaments forming shot-lived spirals that quickly uncoil. In this regime, semi-flexible filaments are actively pushed against each other enhancing shape fluctuations which prevents formation 
of large-scale flow patterns.  

The onset of the instability is controlled by the flexure number, $\mathfrak{F}$. We find a similar threshold, $\mathfrak{F}\approx 10^3$, as in the single-filament case~\cite{riseleholder-15} 
(grey-shaded region in FIG.~\ref{fig:phase-diagram}). For $Pe\gtrsim10^3$, the coherently moving bundles dissolve and the system regains a nearly uniform density. The structure of this state 
is markedly different from the low activity case. Most filaments curl into long-lived \emph{spirals} (FIG.~\ref{fig:phase}d). While the head bead of
the filament is always in the centre, there is no preferred direction of the rotation, i.e., there is no global chirality. Centres of spirals move diffusively (FIG.~\ref{fig:struct-dynamic}a). If the system 
density is reduced, the state resembles a weakly interacting gas of rotating spirals~\cite{supporting}. It is interesting to ask why the system prefers this unusual spiralling state, as opposed to, 
e.g., a configuration where perfectly aligned fully stretched filaments flow parallel to each other, which would be preferential energetically. We argue that it is ultimately connected to the conformational 
entropy of the chains: Straight filament configurations are entropically costly and spontaneous shape fluctuations would affect the entire flow. On the other hand, the entropically favourable coiled 
conformations are not compatible with the self-propulsion which prefers coherent motion with a constant speed $\approx f_p/\gamma$. Therefore, the system balances these two competing effect by 
selecting filament conformations that trap most activity into circular motion. This mechanism has no analogue in systems of simple structureless active agents and owes its existence solely
 to the extended nature of the filaments, as can also be seen by reducing $N$, which leads to the disappearance of the spiralling state.

In order to characterise the spiralling state, we compute the average angular velocity of filaments around their centres of mass. The instantaneous  angular velocity of filament $j$ is 
\begin{equation}
\boldsymbol{\omega}_{j}\left(t\right)=\frac{1}{N}\sum_{i=1}^N\frac{\left(\mathbf{r}_{i}(t)-\mathbf{r}_{j,cm}(t)\right)\times\left(\mathbf{v}_{i}(t) -\mathbf{v}_{j,cm}(t)\right)}{\left|\mathbf{r}_{i}(t)-\mathbf{r}_{j,cm}(t)\right|^2},
\end{equation}
where $\mathbf{r}_i(t)$ and $\mathbf{v}_i(t)$ are, respectively, the position and velocity of the bead $i$ at time $t$. Similarly, $\mathbf{r}_{j,cm}(t)$ and $\mathbf{v}_{j,cm}(t)$ are the instantaneous 
position and velocity of the centre of mass of filament $j$, respectively. The average magnitude of the angular velocity per filament is then 
\begin{equation}
 \omega=\frac{1}{M}\sum_{j=1}^M\left|\frac{1}{\tau_{m}}\sum_t\boldsymbol{\omega}_j(t)\right|, \label{eq:omega}
 \end{equation}
 where $\tau_{m}$ is the measurement time. We note that the order in which averages are taken is important to properly capture rotations of individual filaments, and also that $\omega$ retains 
 a small but finite plateau at low $Pe$. In FIG.~\ref{fig:struct-dynamic}d we show that $\omega$ increases sharply from its plateau value as a function of P{\'e}clet number as the buckling threshold 
 is crossed, confirming that individual spirals rotate. The rate of the increase of $\omega$ with $Pe$ grows with filaments stiffness, which is not surprising as softer filaments are easier to bend.

One of the identifying features of self-propelled filaments is the reduction of the effective bending stiffness manifested as the decrease of the effective persistence length, $\xi_p$. The persistence 
length, $\xi_p^a$, is determined by fitting the tangent-tangent correlation function $\langle\mathbf{t}(s)\cdot\mathbf{t}(0)\rangle$ to $\exp\left(-s/\xi_p\right)$, where $s$ measures the position 
along the contour. In FIG.~\ref{fig:struct-dynamic}c we show the ratio of the persistence length in the presence of activity and its passive, thermal value, $\xi_p^a/\xi_p$ as a function of $Pe$ for 
a range of bare bending stiffnesses, $\kappa$. For low values of $Pe$,  $\xi_p^a/\xi_p$ remains close to $1$, indicating that the filament stiffness is not affected by activity. However, as the 
activity increases, one observes a rapid decrease of the persistence length. This activity-driven reduction of $\xi_p^a$ is accompanied by a drop in the radius of gyration (FIG.~\ref{fig:struct-dynamic}c, inset), 
indicating that the system transitions into a coiled state.

Finally, we explore how orientational order of semi-flexible filaments emerges from self-propulsion. Steric repulsion leads to local alignment which does not depend on the filament direction and, therefore, 
has nematic symmetry. On the other hand, self-propulsion introduces directionality, i.e., filament polarity. Therefore, we compute both polar and nematic local order parameter of the filament tangent vector
 $\mathbf{t}_i \equiv ( \mathbf{r}_i-\mathbf{r}_{i-1})/| \mathbf{r}_i-\mathbf{r}_{i-1}|$: Let $\theta_{ij}$ be the angle between tangent vectors $\mathbf{t}_i$ and $\mathbf{t}_j$, belonging to the same or different filaments. Then the order parameter is 
\begin{equation}
S_m=\langle\cos\left(m\theta_{ij}\right)\rangle,
\label{eq:op}
\end{equation}
with $m=1$ for polar, $m=2$ for nematic and where the average $\langle \cdot \rangle$ is over all pairs within a cut-off distance, $5\sigma$.

\begin{figure}[tb] 
\centering
\includegraphics[width=\columnwidth]{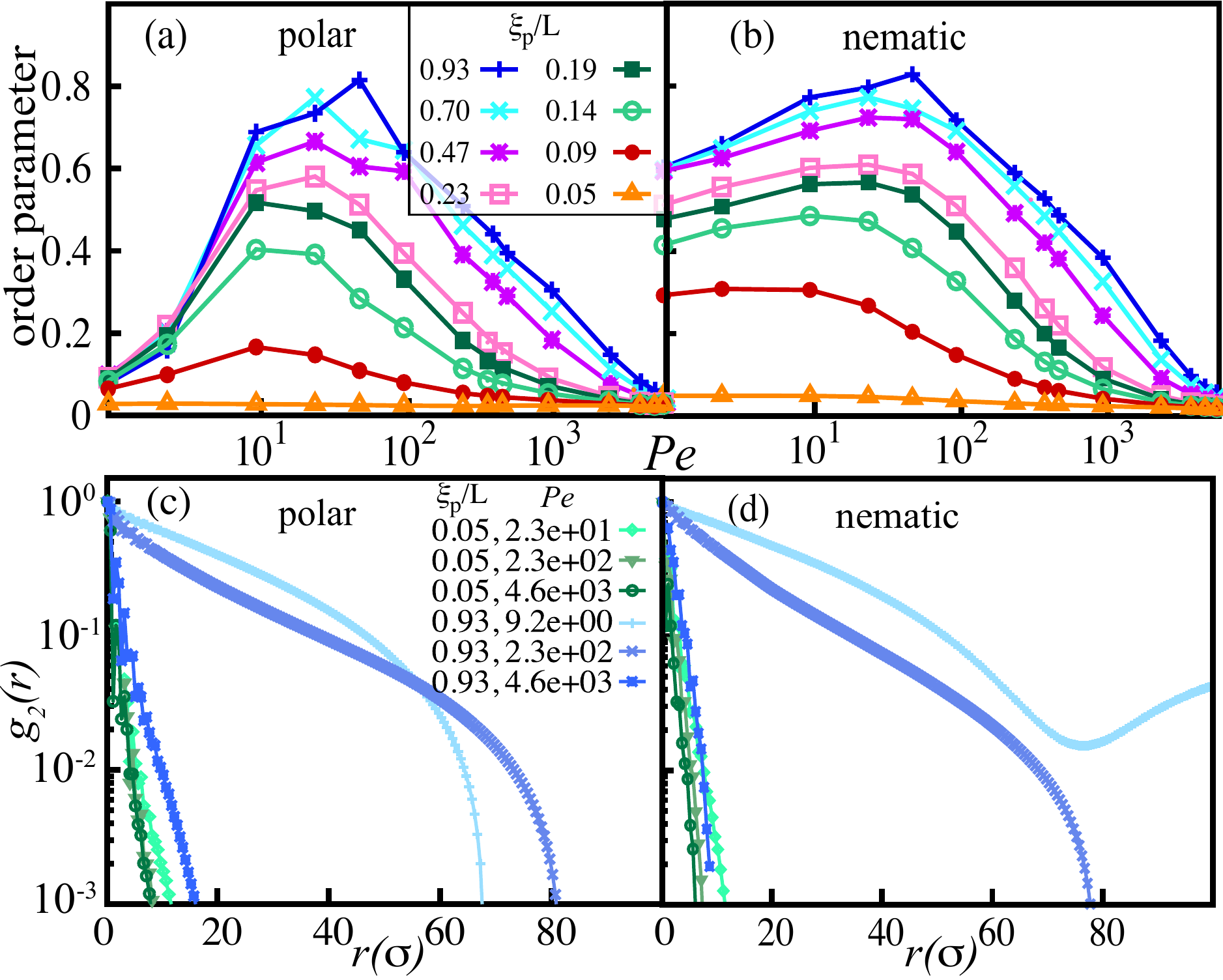} 
\caption{(Colour online) Local polar (a) and nematic (b) order parameter defined in Eq.~(\ref{eq:op}) as a function of activity for a range of bare filament stiffnesses for $N=25$. Spatial correlation 
function $g_{m,2}(r)$ defined in Eq.~(\ref{eq:correl}) for the polar (c) and nematic (d) order parameter. Note that the upturn in (d) is an artefact of averaging for large $r$. }
\label{fig:gr-nem-polar-polar-head}
\end{figure}

In FIG.~\ref{fig:gr-nem-polar-polar-head}a we show the local polar order parameter $S_1$ as a function of $Pe$. For low $Pe$, the effects of activity are very weak and there is essentially no polar 
ordering. As the activity increases, filaments align their propulsion directions and start to flow as a bundle, leading to a boost in $S_1$ for the intermediate values of $Pe$. At large activities, the 
system collapses into the spiralling phase, and the order disappears. We observe similar behaviour for the local nematic order $S_2$ (FIG.~\ref{fig:gr-nem-polar-polar-head}b), however, at low activity the system exhibits substantial local nematic order, consistent with passive polymer melts~\cite{doi1988theory}.
 
In order to probe the extent of the local order in FIG.~\ref{fig:gr-nem-polar-polar-head}c\&d we show the spatial pair-correlations of both polar and nematic order parameters of tangent vectors $\mathbf{t}_i$ and $\mathbf{t}_j$ separated by a distance $r$:  
\begin{equation} 
g_{m,2}(r)=\frac{\langle\sum_{i,j}\delta\left(r-\left|\mathbf{r}_i-\mathbf{r}_j\right|\right)\cos\left(m\theta_{ij}\right)\rangle}{\langle \sum_{i,j} \delta\left(r-\left|\mathbf{r}_i-\mathbf{r}_j\right|\right) \rangle}.
\label{eq:correl}
\end{equation}
In all regimes, we observed an exponential decay of $g_{m,2}$. There is, however, a clear difference between flexible and stiff filaments in terms of the extent of spatial correlations. For flexible filaments 
the order is indeed local and $g_{m,2}$ rapidly drops to zero at distances $\sim10\sigma$. However, for stiff filaments, both polar and nematic order persist over much longer distances, reaching 
a fifth of the system size. In the high activity regime, coiling into spirals, however, completely destroys the order even for the stiffest filaments.

\section{Summary and Conclusions}
\label{sec:summary}
In this paper we used Langevin dynamics simulations of an agent-based model to study collective behaviour of a dense suspension of semi-flexible filaments subject to an active force acting in the 
direction of the filaments' contours. We took steric effects into account, preventing any intersections between filaments. Furthermore, we assumed the dry limit, 
i.e. solvent-mediated hydrodynamic interactions were ignored, an assumption that is justifiable when modelling motility assay experiments. We mapped a detailed non-equilibrium phase diagram 
as a function of activity, measured in terms of the P{\'e}clet number $Pe$, and filament stiffness, measured as the ratio of the passive persistence length to the filament length, $\xi_p/L$, for several 
filament lengths. 

The intricate interplay between activity and conformational changes leads to rich collective behaviour in this system. In particular, at low activity, we found a slowly flowing melt-like state, with 
prominent half-integer topological defects. Those defects, however, are moving very slowly, consistent with other studies of self-propelled nematics.

At intermediate values of activity we observed phase segregation into a state of aligned bundles akin to a MIPS phase. Finally, at very high activity, this state disappears and we observed a peculiar 
spiralling state characterised by filaments predominantly curling themselves into rotating spirals. In this state, the density is again uniform with no global flow. This suggests a mechanism by which the 
system expels activity in part by changing the conformation of the filaments. To the best of our knowledge, this effect has no analogue in systems of active agents with no internal structure.

It would be interesting to fully characterise the nature of the flowing melt phase, where one observes spontaneous formation of half-integer topological defects. Preliminary results 
for a related model where activity acts on pairs of polymers show extensile active dynamics at the time scale of polymer motion, and no polar components to the flow, hinting at a 
fundamental role of the local symmetry of active driving.


The model studied in this paper is too simplified to quantitatively describe a specific experiment. However, it captures several generic collective active patterns and we hope that results presented 
here will motivate further research in the effects of activity on collective behaviour of dense filamentous systems. In particular, experiments on motility assays could provide realisations of some 
of the phases reported here, especially in the the region of phase space where filaments are sufficiently stiff. 

Another interesting system for which our results could be of value for are experiments on clustering of myxobacteria during the vegetative phase~\cite{reichenbach1965swarm}. Myxobacteria are known to form 
a rich variety of collective patterns despite the absence of any long range interactions, such as chemotaxis~\cite{kaiser2003coupling}. This is further substantiated by recent experiments~\cite{drescher2011fluid} 
that have shown that hydrodynamic effects play a negligible role in the collective dynamics of bacterial suspensions.
 

\begin{acknowledgments}
RS acknowledges the support from UK EPSRC (EP/M009599/1) and SH acknowledges support from UK BBSRC (BB/N009150/1). We thank A.~Das, L.~Giomi, A.~Maitra, M.~C.~Marchetti, S.~Saha 
and R.~Selinger for many useful discussions. KRP would like to thank A.~M.~Strydom, University of Johannesburg for hospitality.
\end{acknowledgments}

\begin{figure}[htb] 
\centering
\includegraphics[width=0.85\columnwidth]{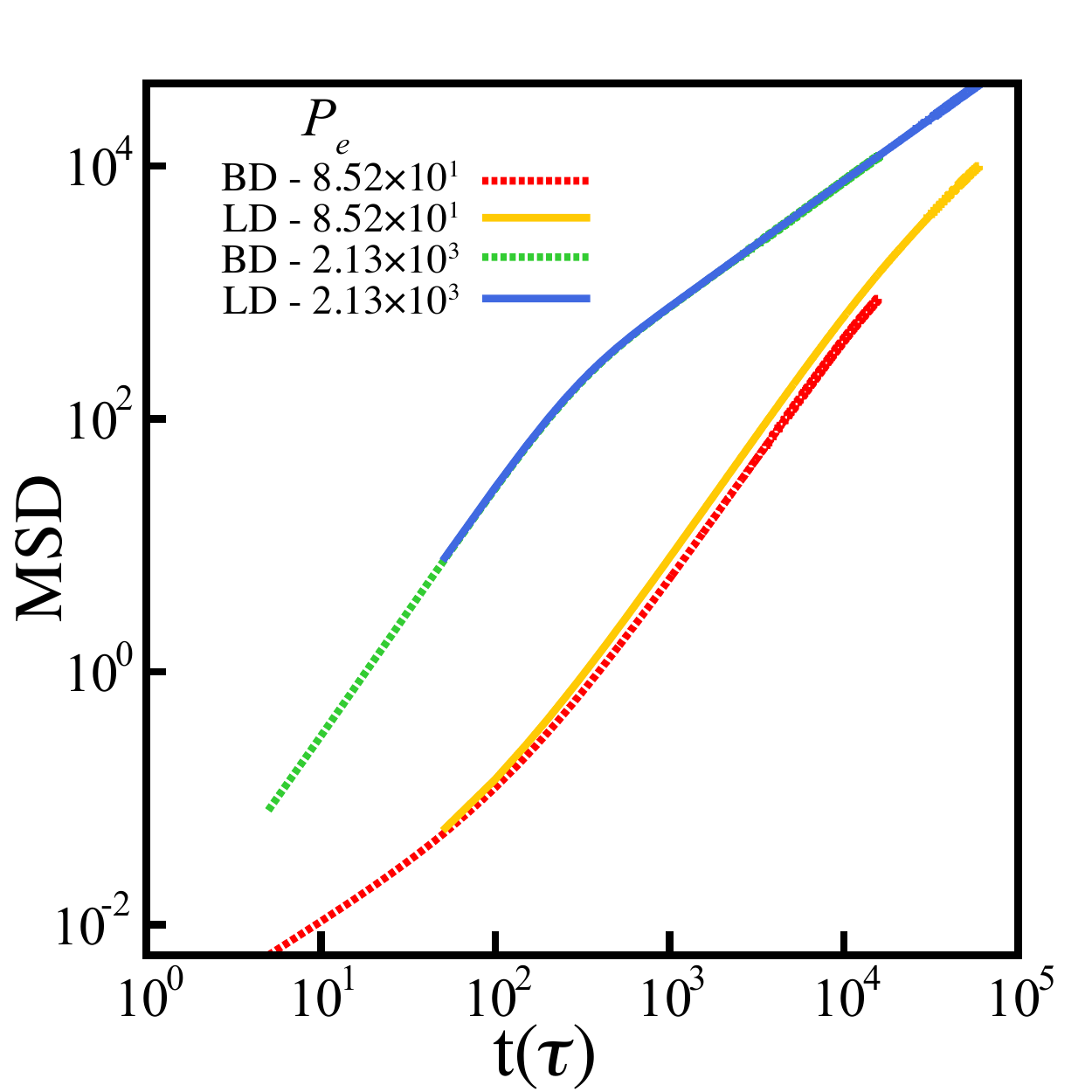} 
\caption{(Colour online) Comparison of the MSD curves for two values of the P{\'e}clet number using Langevin (solid lines) and Brownian (dashed lines) dynamics simulations. For large activity, 
at long times, the MSD curves are indistinguishable. For smaller values of $Pe$, there is a small difference between the two, but we attribute it to very slow dynamics in this regime: it is likely that 
the steady state has not been reached during the duration of the simulation.  }
\label{fig:MSD_comparison}
\end{figure}

\appendix
\section{Langevin vs.~Brownian dynamics} 
\label{app:langevin_vs_brownian}
In this Appendix we show a comparison of the MSD curves for centre of mass of filaments in several different regimes produced using both Langevin and Brownian dynamics simulations. Brownian 
dynamics simulations were performed using LAMMPS extended by a publicly available Brownian Dynamics ``fix''~\cite{lammps_bd}. In FIG.~\ref{fig:MSD_comparison} we show plots of the MSD curves
for two different values of the P{\'e}clet number, $Pe$. From these plots it is evident that for sufficiently long times, MSD curves obtained by those two different methods coincide in the high-activity regime 
and are very close to each other for low values of $Pe$. We attribute the discrepancy between the two in the low $Pe$ regime to very slow collective dynamics. These results show that on at long time scales, 
there should be no qualitative difference between the two methods. We note, however, that Brownian dynamics simulations become unstable when the simulation time step is increased beyond $10^{-4}$, 
which is $10$ times smaller than the time step used in the Langevin dynamics simulations. This means that the using Langevin dynamics simulations one can effectively reach $10$ longer times compared 
to what is possible with Brownian dynamics simulations using the same computational resources. A detailed discussion of the numerical stability of different methods for integrating equations of motion can 
be found, e.g.~in Ref.~\cite{leimkuhler2015molecular}.

\section{Identifying topological defects}
\label{app:defects}
\begin{figure}[htb] 
\centering
\includegraphics[width=0.95\columnwidth]{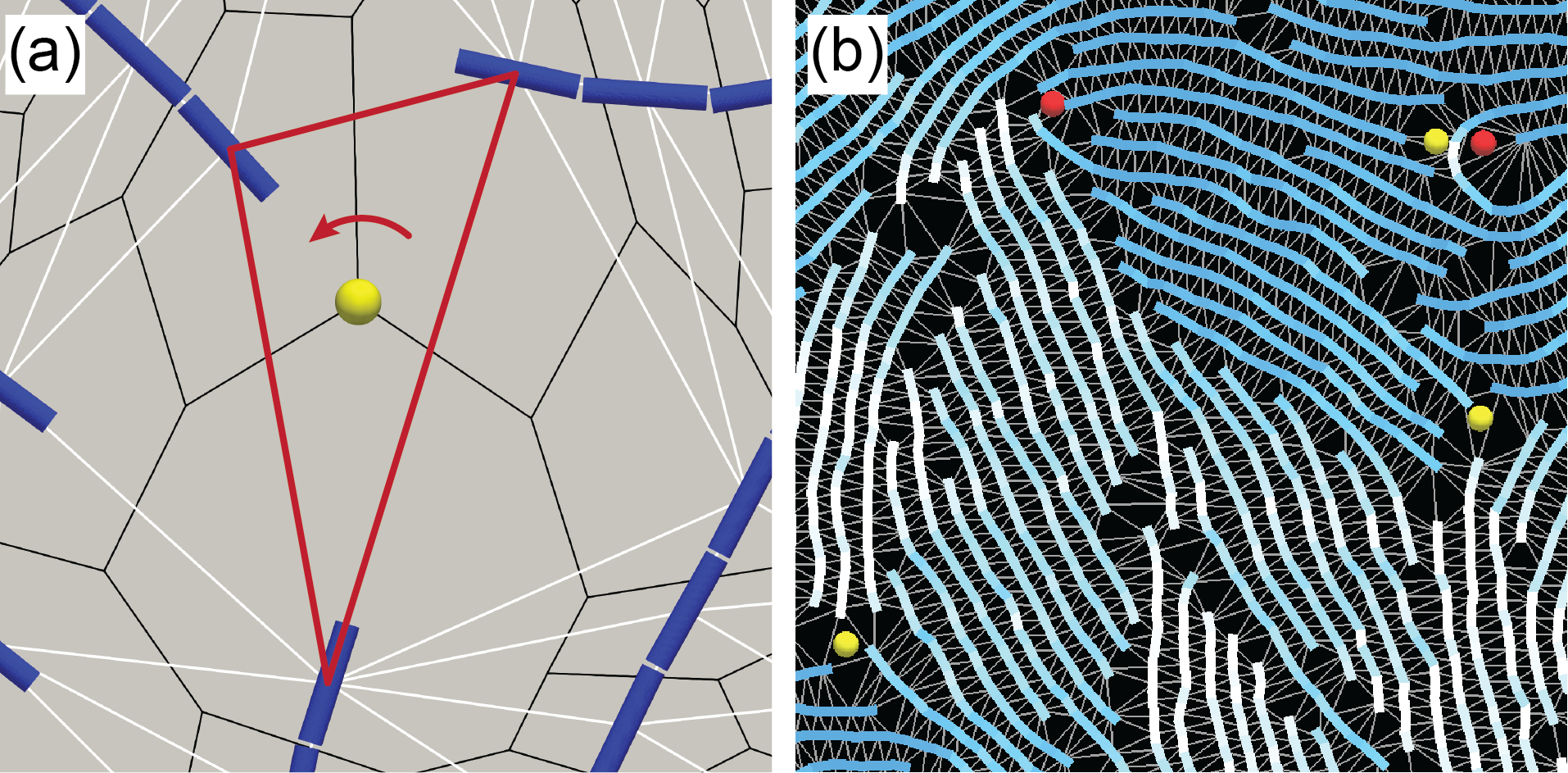} 
\caption{(Colour online) Identification of topological defects. (a) Headless unit-length vectors pointing along the local tangent to the filament contour are placed at the vertices of a Delaunay triangulation (white lines).
We then loop counterclockwise (red arrow) over the vertices of each triangle (marked in red) and compute a signed sum of the angles the headless vectors make with the $x-$axis. If this sum exceeds 
$\pm\pi$ a defect of appropriate topological charge is places inside that triangle. Defects (yellow ball) are located at the vertices of the Voronoi diagram (black lines) dual to the Delaunay 
triangulation. (b) Application of this algorithm to a snapshot of the simulation. For visualisation purposes, we zoomed in on only a couple of defects.   }
\label{fig:defects_algorthm}
\end{figure}

In this Appendix we briefly outline the method used to identify topological defects. We assign to each bead of each filament a headless (nematic) unit-length vector $\mathbf{t}_i$ pointing along the 
local tangent to the contour. The direction of the tangent is determined as an average of the directions along the two bonds a bead belongs to. We use the positions of the beads as vertices of a Delaunay 
triangulation, taking into account periodic boundary conditions. We then loop over the vertices of each triangle in the counterclockwise direction and sum signed angles between vectors $\mathbf{t}_i$ and the  
$x-$axis. Due to the nematic symmetry these angles are between $-\pi/2$ and $\pi/2$. If the sum exceeds $\pm n \pi$ (where $n$ is an integer), we assign a defect of charge $\pm n/2$ to the centre of 
that triangle. This method and its application are illustrated in Fig.~\ref{fig:defects_algorthm}. 

\begin{thebibliography}{67}%
\makeatletter
\providecommand \@ifxundefined [1]{%
 \@ifx{#1\undefined}
}%
\providecommand \@ifnum [1]{%
 \ifnum #1\expandafter \@firstoftwo
 \else \expandafter \@secondoftwo
 \fi
}%
\providecommand \@ifx [1]{%
 \ifx #1\expandafter \@firstoftwo
 \else \expandafter \@secondoftwo
 \fi
}%
\providecommand \natexlab [1]{#1}%
\providecommand \enquote  [1]{``#1''}%
\providecommand \bibnamefont  [1]{#1}%
\providecommand \bibfnamefont [1]{#1}%
\providecommand \citenamefont [1]{#1}%
\providecommand \href@noop [0]{\@secondoftwo}%
\providecommand \href [0]{\begingroup \@sanitize@url \@href}%
\providecommand \@href[1]{\@@startlink{#1}\@@href}%
\providecommand \@@href[1]{\endgroup#1\@@endlink}%
\providecommand \@sanitize@url [0]{\catcode `\\12\catcode `\$12\catcode
  `\&12\catcode `\#12\catcode `\^12\catcode `\_12\catcode `\%12\relax}%
\providecommand \@@startlink[1]{}%
\providecommand \@@endlink[0]{}%
\providecommand \url  [0]{\begingroup\@sanitize@url \@url }%
\providecommand \@url [1]{\endgroup\@href {#1}{\urlprefix }}%
\providecommand \urlprefix  [0]{URL }%
\providecommand \Eprint [0]{\href }%
\providecommand \doibase [0]{http://dx.doi.org/}%
\providecommand \selectlanguage [0]{\@gobble}%
\providecommand \bibinfo  [0]{\@secondoftwo}%
\providecommand \bibfield  [0]{\@secondoftwo}%
\providecommand \translation [1]{[#1]}%
\providecommand \BibitemOpen [0]{}%
\providecommand \bibitemStop [0]{}%
\providecommand \bibitemNoStop [0]{.\EOS\space}%
\providecommand \EOS [0]{\spacefactor3000\relax}%
\providecommand \BibitemShut  [1]{\csname bibitem#1\endcsname}%
\let\auto@bib@innerbib\@empty
\bibitem [{\citenamefont {Alberts}\ \emph {et~al.}(2008)\citenamefont
  {Alberts}, \citenamefont {Johnson}, \citenamefont {Walter}, \citenamefont
  {Lewis}, \citenamefont {Raff},\ and\ \citenamefont
  {Roberts}}]{alberts-thecell-08}%
  \BibitemOpen
  \bibfield  {author} {\bibinfo {author} {\bibfnamefont {B.}~\bibnamefont
  {Alberts}}, \bibinfo {author} {\bibfnamefont {A.}~\bibnamefont {Johnson}},
  \bibinfo {author} {\bibfnamefont {P.}~\bibnamefont {Walter}}, \bibinfo
  {author} {\bibfnamefont {J.}~\bibnamefont {Lewis}}, \bibinfo {author}
  {\bibfnamefont {M.}~\bibnamefont {Raff}}, \ and\ \bibinfo {author}
  {\bibfnamefont {K.}~\bibnamefont {Roberts}},\ }\href@noop {} {\emph {\bibinfo
  {title} {Molecular biology of the Cell}}}\ (\bibinfo  {publisher} {Taylor and
  Francis Inc},\ \bibinfo {address} {CT, US},\ \bibinfo {year}
  {2008})\BibitemShut {NoStop}%
\bibitem [{\citenamefont {Fletcher}\ and\ \citenamefont
  {Mullins}(2010)}]{fletcher2010cell}%
  \BibitemOpen
  \bibfield  {author} {\bibinfo {author} {\bibfnamefont {D.~A.}\ \bibnamefont
  {Fletcher}}\ and\ \bibinfo {author} {\bibfnamefont {R.~D.}\ \bibnamefont
  {Mullins}},\ }\href@noop {} {\bibfield  {journal} {\bibinfo  {journal}
  {Nature}\ }\textbf {\bibinfo {volume} {463}},\ \bibinfo {pages} {485}
  (\bibinfo {year} {2010})}\BibitemShut {NoStop}%
\bibitem [{\citenamefont {Huber}\ \emph {et~al.}(2013)\citenamefont {Huber},
  \citenamefont {Schnau{\ss}}, \citenamefont {R{\"o}nicke}, \citenamefont
  {Rauch}, \citenamefont {M{\"u}ller}, \citenamefont {F{\"u}tterer},\ and\
  \citenamefont {K{\"a}s}}]{huber2013emergent}%
  \BibitemOpen
  \bibfield  {author} {\bibinfo {author} {\bibfnamefont {F.}~\bibnamefont
  {Huber}}, \bibinfo {author} {\bibfnamefont {J.}~\bibnamefont {Schnau{\ss}}},
  \bibinfo {author} {\bibfnamefont {S.}~\bibnamefont {R{\"o}nicke}}, \bibinfo
  {author} {\bibfnamefont {P.}~\bibnamefont {Rauch}}, \bibinfo {author}
  {\bibfnamefont {K.}~\bibnamefont {M{\"u}ller}}, \bibinfo {author}
  {\bibfnamefont {C.}~\bibnamefont {F{\"u}tterer}}, \ and\ \bibinfo {author}
  {\bibfnamefont {J.}~\bibnamefont {K{\"a}s}},\ }\href@noop {} {\bibfield
  {journal} {\bibinfo  {journal} {Adv. Phys.}\ }\textbf {\bibinfo
  {volume} {62}},\ \bibinfo {pages} {1} (\bibinfo {year} {2013})}\BibitemShut
  {NoStop}%
\bibitem [{\citenamefont {Warrick}\ and\ \citenamefont
  {Spudich}(1987)}]{warrick1987myosin}%
  \BibitemOpen
  \bibfield  {author} {\bibinfo {author} {\bibfnamefont {H.~M.}\ \bibnamefont
  {Warrick}}\ and\ \bibinfo {author} {\bibfnamefont {J.~A.}\ \bibnamefont
  {Spudich}},\ }\href@noop {} {\bibfield  {journal} {\bibinfo  {journal}
  {Ann. Rev. Cell Bio.}\ }\textbf {\bibinfo {volume} {3}},\ \bibinfo
  {pages} {379} (\bibinfo {year} {1987})}\BibitemShut {NoStop}%
\bibitem [{\citenamefont {Ridley}\ \emph {et~al.}(2003)\citenamefont {Ridley},
  \citenamefont {Schwartz}, \citenamefont {Burridge}, \citenamefont {Firtel},
  \citenamefont {Ginsberg}, \citenamefont {Borisy}, \citenamefont {Parsons},\
  and\ \citenamefont {Horwitz}}]{ridley2003cell}%
  \BibitemOpen
  \bibfield  {author} {\bibinfo {author} {\bibfnamefont {A.~J.}\ \bibnamefont
  {Ridley}}, \bibinfo {author} {\bibfnamefont {M.~A.}\ \bibnamefont
  {Schwartz}}, \bibinfo {author} {\bibfnamefont {K.}~\bibnamefont {Burridge}},
  \bibinfo {author} {\bibfnamefont {R.~A.}\ \bibnamefont {Firtel}}, \bibinfo
  {author} {\bibfnamefont {M.~H.}\ \bibnamefont {Ginsberg}}, \bibinfo {author}
  {\bibfnamefont {G.}~\bibnamefont {Borisy}}, \bibinfo {author} {\bibfnamefont
  {J.~T.}\ \bibnamefont {Parsons}}, \ and\ \bibinfo {author} {\bibfnamefont
  {A.~R.}\ \bibnamefont {Horwitz}},\ }\href@noop {} {\bibfield  {journal}
  {\bibinfo  {journal} {Science}\ }\textbf {\bibinfo {volume} {302}},\ \bibinfo
  {pages} {1704} (\bibinfo {year} {2003})}\BibitemShut {NoStop}%
\bibitem [{\citenamefont {Sheetz}\ and\ \citenamefont
  {Spudich}(1983)}]{sheetz1983movement}%
  \BibitemOpen
  \bibfield  {author} {\bibinfo {author} {\bibfnamefont {M.~P.}\ \bibnamefont
  {Sheetz}}\ and\ \bibinfo {author} {\bibfnamefont {J.~A.}\ \bibnamefont
  {Spudich}},\ }\href@noop {} {\bibfield  {journal} {\bibinfo  {journal}
  {Nature}\ }\textbf {\bibinfo {volume} {303}},\ \bibinfo {pages} {31}
  (\bibinfo {year} {1983})}\BibitemShut {NoStop}%
\bibitem [{\citenamefont {Sheetz}\ \emph {et~al.}(1984)\citenamefont {Sheetz},
  \citenamefont {Chasan},\ and\ \citenamefont {Spudich}}]{sheetz1984atp}%
  \BibitemOpen
  \bibfield  {author} {\bibinfo {author} {\bibfnamefont {M.~P.}\ \bibnamefont
  {Sheetz}}, \bibinfo {author} {\bibfnamefont {R.}~\bibnamefont {Chasan}}, \
  and\ \bibinfo {author} {\bibfnamefont {J.~A.}\ \bibnamefont {Spudich}},\
  }\href@noop {} {\bibfield  {journal} {\bibinfo  {journal} {J. Cell Biol}\
  }\textbf {\bibinfo {volume} {99}},\ \bibinfo {pages} {1867} (\bibinfo {year}
  {1984})}\BibitemShut {NoStop}%
\bibitem [{\citenamefont {N{\'e}d{\'e}lec}\ \emph {et~al.}(1997)\citenamefont
  {N{\'e}d{\'e}lec}, \citenamefont {Surrey}, \citenamefont {Maggs},\ and\
  \citenamefont {Leibler}}]{ndlec1997self}%
  \BibitemOpen
  \bibfield  {author} {\bibinfo {author} {\bibfnamefont {F.}~\bibnamefont
  {N{\'e}d{\'e}lec}}, \bibinfo {author} {\bibfnamefont {T.}~\bibnamefont
  {Surrey}}, \bibinfo {author} {\bibfnamefont {A.~C.}\ \bibnamefont {Maggs}}, \
  and\ \bibinfo {author} {\bibfnamefont {S.}~\bibnamefont {Leibler}},\
  }\href@noop {} {\bibfield  {journal} {\bibinfo  {journal} {Nature}\ }\textbf
  {\bibinfo {volume} {389}},\ \bibinfo {pages} {305} (\bibinfo {year}
  {1997})}\BibitemShut {NoStop}%
\bibitem [{\citenamefont {Surrey}\ \emph {et~al.}(2001)\citenamefont {Surrey},
  \citenamefont {N{\'e}d{\'e}lec}, \citenamefont {Leibler},\ and\ \citenamefont
  {Karsenti}}]{surrey2001physical}%
  \BibitemOpen
  \bibfield  {author} {\bibinfo {author} {\bibfnamefont {T.}~\bibnamefont
  {Surrey}}, \bibinfo {author} {\bibfnamefont {F.}~\bibnamefont
  {N{\'e}d{\'e}lec}}, \bibinfo {author} {\bibfnamefont {S.}~\bibnamefont
  {Leibler}}, \ and\ \bibinfo {author} {\bibfnamefont {E.}~\bibnamefont
  {Karsenti}},\ }\href@noop {} {\bibfield  {journal} {\bibinfo  {journal}
  {Science}\ }\textbf {\bibinfo {volume} {292}},\ \bibinfo {pages} {1167}
  (\bibinfo {year} {2001})}\BibitemShut {NoStop}%
\bibitem [{\citenamefont {e~Silva}\ \emph {et~al.}(2011)\citenamefont
  {e~Silva}, \citenamefont {Depken}, \citenamefont {Stuhrmann}, \citenamefont
  {Korsten}, \citenamefont {MacKintosh},\ and\ \citenamefont
  {Koenderink}}]{e2011active}%
  \BibitemOpen
  \bibfield  {author} {\bibinfo {author} {\bibfnamefont {M.~S.}\ \bibnamefont
  {e~Silva}}, \bibinfo {author} {\bibfnamefont {M.}~\bibnamefont {Depken}},
  \bibinfo {author} {\bibfnamefont {B.}~\bibnamefont {Stuhrmann}}, \bibinfo
  {author} {\bibfnamefont {M.}~\bibnamefont {Korsten}}, \bibinfo {author}
  {\bibfnamefont {F.~C.}\ \bibnamefont {MacKintosh}}, \ and\ \bibinfo {author}
  {\bibfnamefont {G.~H.}\ \bibnamefont {Koenderink}},\ }\href@noop {}
  {\bibfield  {journal} {\bibinfo  {journal} {Proceedings of the National
  Academy of Sciences}\ }\textbf {\bibinfo {volume} {108}},\ \bibinfo {pages}
  {9408} (\bibinfo {year} {2011})}\BibitemShut {NoStop}%
\bibitem [{\citenamefont {K{\"o}ster}\ \emph {et~al.}(2016)\citenamefont
  {K{\"o}ster}, \citenamefont {Husain}, \citenamefont {Iljazi}, \citenamefont
  {Bhat}, \citenamefont {Bieling}, \citenamefont {Mullins}, \citenamefont
  {Rao},\ and\ \citenamefont {Mayor}}]{koster2016actomyosin}%
  \BibitemOpen
  \bibfield  {author} {\bibinfo {author} {\bibfnamefont {D.~V.}\ \bibnamefont
  {K{\"o}ster}}, \bibinfo {author} {\bibfnamefont {K.}~\bibnamefont {Husain}},
  \bibinfo {author} {\bibfnamefont {E.}~\bibnamefont {Iljazi}}, \bibinfo
  {author} {\bibfnamefont {A.}~\bibnamefont {Bhat}}, \bibinfo {author}
  {\bibfnamefont {P.}~\bibnamefont {Bieling}}, \bibinfo {author} {\bibfnamefont
  {R.~D.}\ \bibnamefont {Mullins}}, \bibinfo {author} {\bibfnamefont
  {M.}~\bibnamefont {Rao}}, \ and\ \bibinfo {author} {\bibfnamefont
  {S.}~\bibnamefont {Mayor}},\ }\href@noop {} {\bibfield  {journal} {\bibinfo
  {journal} {Pro. Nat. Acad. Sci. USA}\ ,\ \bibinfo {pages} {201514030}}
  (\bibinfo {year} {2016})}\BibitemShut {NoStop}%
\bibitem [{\citenamefont {Linsmeier}\ \emph {et~al.}(2016)\citenamefont
  {Linsmeier}, \citenamefont {Banerjee}, \citenamefont {Oakes}, \citenamefont
  {Jung}, \citenamefont {Kim},\ and\ \citenamefont
  {Murrell}}]{linsmeier2016disordered}%
  \BibitemOpen
  \bibfield  {author} {\bibinfo {author} {\bibfnamefont {I.}~\bibnamefont
  {Linsmeier}}, \bibinfo {author} {\bibfnamefont {S.}~\bibnamefont {Banerjee}},
  \bibinfo {author} {\bibfnamefont {P.~W.}\ \bibnamefont {Oakes}}, \bibinfo
  {author} {\bibfnamefont {W.}~\bibnamefont {Jung}}, \bibinfo {author}
  {\bibfnamefont {T.}~\bibnamefont {Kim}}, \ and\ \bibinfo {author}
  {\bibfnamefont {M.~P.}\ \bibnamefont {Murrell}},\ }\href@noop {} {\bibfield
  {journal} {\bibinfo  {journal} {Nat.~Commun.}\ }\textbf {\bibinfo
  {volume} {7}},\ \bibinfo {pages} {12615} (\bibinfo {year}
  {2016})}\BibitemShut {NoStop}%
\bibitem [{\citenamefont {Kruse}\ \emph {et~al.}(2004)\citenamefont {Kruse},
  \citenamefont {Joanny}, \citenamefont {J{\"u}licher}, \citenamefont {Prost},\
  and\ \citenamefont {Sekimoto}}]{kruse2004asters}%
  \BibitemOpen
  \bibfield  {author} {\bibinfo {author} {\bibfnamefont {K.}~\bibnamefont
  {Kruse}}, \bibinfo {author} {\bibfnamefont {J.-F.}\ \bibnamefont {Joanny}},
  \bibinfo {author} {\bibfnamefont {F.}~\bibnamefont {J{\"u}licher}}, \bibinfo
  {author} {\bibfnamefont {J.}~\bibnamefont {Prost}}, \ and\ \bibinfo {author}
  {\bibfnamefont {K.}~\bibnamefont {Sekimoto}},\ }\href@noop {} {\bibfield
  {journal} {\bibinfo  {journal} {Phys. Rev. Lett.}\ }\textbf {\bibinfo
  {volume} {92}},\ \bibinfo {pages} {078101} (\bibinfo {year}
  {2004})}\BibitemShut {NoStop}%
\bibitem [{\citenamefont {Ramaswamy}(2010)}]{sriram-10}%
  \BibitemOpen
  \bibfield  {author} {\bibinfo {author} {\bibfnamefont {S.}~\bibnamefont
  {Ramaswamy}},\ }\href@noop {} {\bibfield  {journal} {\bibinfo  {journal}
  {Annu. Rev. Condens. Matter Phys.}\ }\textbf {\bibinfo {volume} {1}},\
  \bibinfo {pages} {323} (\bibinfo {year} {2010})}\BibitemShut {NoStop}%
\bibitem [{\citenamefont {Marchetti}\ \emph {et~al.}(2013)\citenamefont
  {Marchetti}, \citenamefont {Joanny}, \citenamefont {Ramaswamy}, \citenamefont
  {Liverpool}, \citenamefont {Prost}, \citenamefont {Rao},\ and\ \citenamefont
  {Simha}}]{marchetti-13}%
  \BibitemOpen
  \bibfield  {author} {\bibinfo {author} {\bibfnamefont {M.}~\bibnamefont
  {Marchetti}}, \bibinfo {author} {\bibfnamefont {J.}~\bibnamefont {Joanny}},
  \bibinfo {author} {\bibfnamefont {S.}~\bibnamefont {Ramaswamy}}, \bibinfo
  {author} {\bibfnamefont {T.}~\bibnamefont {Liverpool}}, \bibinfo {author}
  {\bibfnamefont {J.}~\bibnamefont {Prost}}, \bibinfo {author} {\bibfnamefont
  {M.}~\bibnamefont {Rao}}, \ and\ \bibinfo {author} {\bibfnamefont {R.~A.}\
  \bibnamefont {Simha}},\ }\href@noop {} {\bibfield  {journal} {\bibinfo
  {journal} {Rev. Mod. Phys.}\ }\textbf {\bibinfo {volume} {85}},\ \bibinfo
  {pages} {1143} (\bibinfo {year} {2013})}\BibitemShut {NoStop}%
\bibitem [{\citenamefont {Murrell}\ \emph {et~al.}(2015)\citenamefont
  {Murrell}, \citenamefont {Oakes}, \citenamefont {Lenz},\ and\ \citenamefont
  {Gardel}}]{murrell2015forcing}%
  \BibitemOpen
  \bibfield  {author} {\bibinfo {author} {\bibfnamefont {M.}~\bibnamefont
  {Murrell}}, \bibinfo {author} {\bibfnamefont {P.~W.}\ \bibnamefont {Oakes}},
  \bibinfo {author} {\bibfnamefont {M.}~\bibnamefont {Lenz}}, \ and\ \bibinfo
  {author} {\bibfnamefont {M.~L.}\ \bibnamefont {Gardel}},\ }\href@noop {}
  {\bibfield  {journal} {\bibinfo  {journal} {Nat. Rev. Mol. Cell
  Bio.}\ }\textbf {\bibinfo {volume} {16}},\ \bibinfo {pages} {486}
  (\bibinfo {year} {2015})}\BibitemShut {NoStop}%
\bibitem [{\citenamefont {Sumino}\ \emph {et~al.}(2012)\citenamefont {Sumino},
  \citenamefont {Nagai}, \citenamefont {Shitaka}, \citenamefont {Tanaka},
  \citenamefont {Yoshikawa}, \citenamefont {Chat{\'e}},\ and\ \citenamefont
  {Oiwa}}]{sumino2012large}%
  \BibitemOpen
  \bibfield  {author} {\bibinfo {author} {\bibfnamefont {Y.}~\bibnamefont
  {Sumino}}, \bibinfo {author} {\bibfnamefont {K.~H.}\ \bibnamefont {Nagai}},
  \bibinfo {author} {\bibfnamefont {Y.}~\bibnamefont {Shitaka}}, \bibinfo
  {author} {\bibfnamefont {D.}~\bibnamefont {Tanaka}}, \bibinfo {author}
  {\bibfnamefont {K.}~\bibnamefont {Yoshikawa}}, \bibinfo {author}
  {\bibfnamefont {H.}~\bibnamefont {Chat{\'e}}}, \ and\ \bibinfo {author}
  {\bibfnamefont {K.}~\bibnamefont {Oiwa}},\ }\href@noop {} {\bibfield
  {journal} {\bibinfo  {journal} {Nature}\ }\textbf {\bibinfo {volume} {483}},\
  \bibinfo {pages} {448} (\bibinfo {year} {2012})}\BibitemShut {NoStop}%
\bibitem [{\citenamefont {Schaller}\ \emph {et~al.}(2010)\citenamefont
  {Schaller}, \citenamefont {Weber}, \citenamefont {Semmrich}, \citenamefont
  {Frey},\ and\ \citenamefont {Bausch}}]{schaller2010polar}%
  \BibitemOpen
  \bibfield  {author} {\bibinfo {author} {\bibfnamefont {V.}~\bibnamefont
  {Schaller}}, \bibinfo {author} {\bibfnamefont {C.}~\bibnamefont {Weber}},
  \bibinfo {author} {\bibfnamefont {C.}~\bibnamefont {Semmrich}}, \bibinfo
  {author} {\bibfnamefont {E.}~\bibnamefont {Frey}}, \ and\ \bibinfo {author}
  {\bibfnamefont {A.~R.}\ \bibnamefont {Bausch}},\ }\href@noop {} {\bibfield
  {journal} {\bibinfo  {journal} {Nature}\ }\textbf {\bibinfo {volume} {467}},\
  \bibinfo {pages} {73} (\bibinfo {year} {2010})}\BibitemShut {NoStop}%
\bibitem [{\citenamefont {Winkler}\ \emph {et~al.}(2017)\citenamefont
  {Winkler}, \citenamefont {Elgeti},\ and\ \citenamefont
  {Gompper}}]{winkler2017active}%
  \BibitemOpen
  \bibfield  {author} {\bibinfo {author} {\bibfnamefont {R.~G.}\ \bibnamefont
  {Winkler}}, \bibinfo {author} {\bibfnamefont {J.}~\bibnamefont {Elgeti}}, \
  and\ \bibinfo {author} {\bibfnamefont {G.}~\bibnamefont {Gompper}},\
  }\href@noop {} {\bibfield  {journal} {\bibinfo  {journal}
  {J.~Phys.~Soc.~Japan}\ }\textbf {\bibinfo {volume} {86}},\ \bibinfo {pages}
  {101014} (\bibinfo {year} {2017})}\BibitemShut {NoStop}%
\bibitem [{\citenamefont {Sanchez}\ \emph {et~al.}(2012)\citenamefont
  {Sanchez}, \citenamefont {Chen}, \citenamefont {DeCamp}, \citenamefont
  {Heymann},\ and\ \citenamefont {Dogic}}]{tsanchez-12}%
  \BibitemOpen
  \bibfield  {author} {\bibinfo {author} {\bibfnamefont {T.}~\bibnamefont
  {Sanchez}}, \bibinfo {author} {\bibfnamefont {D.~T.~N.}\ \bibnamefont
  {Chen}}, \bibinfo {author} {\bibfnamefont {S.~J.}\ \bibnamefont {DeCamp}},
  \bibinfo {author} {\bibfnamefont {M.}~\bibnamefont {Heymann}}, \ and\
  \bibinfo {author} {\bibfnamefont {Z.}~\bibnamefont {Dogic}},\ }\href@noop {}
  {\bibfield  {journal} {\bibinfo  {journal} {Nature}\ }\textbf {\bibinfo
  {volume} {491}},\ \bibinfo {pages} {431} (\bibinfo {year}
  {2012})}\BibitemShut {NoStop}%
\bibitem [{\citenamefont {Keber}\ \emph {et~al.}(2014)\citenamefont {Keber},
  \citenamefont {Loiseau}, \citenamefont {Sanchez}, \citenamefont {DeCamp},
  \citenamefont {Giomi}, \citenamefont {Bowick}, \citenamefont {Marchetti},
  \citenamefont {Dogic},\ and\ \citenamefont {Bausch}}]{keber2014topology}%
  \BibitemOpen
  \bibfield  {author} {\bibinfo {author} {\bibfnamefont {F.~C.}\ \bibnamefont
  {Keber}}, \bibinfo {author} {\bibfnamefont {E.}~\bibnamefont {Loiseau}},
  \bibinfo {author} {\bibfnamefont {T.}~\bibnamefont {Sanchez}}, \bibinfo
  {author} {\bibfnamefont {S.~J.}\ \bibnamefont {DeCamp}}, \bibinfo {author}
  {\bibfnamefont {L.}~\bibnamefont {Giomi}}, \bibinfo {author} {\bibfnamefont
  {M.~J.}\ \bibnamefont {Bowick}}, \bibinfo {author} {\bibfnamefont {M.~C.}\
  \bibnamefont {Marchetti}}, \bibinfo {author} {\bibfnamefont {Z.}~\bibnamefont
  {Dogic}}, \ and\ \bibinfo {author} {\bibfnamefont {A.~R.}\ \bibnamefont
  {Bausch}},\ }\href@noop {} {\bibfield  {journal} {\bibinfo  {journal}
  {Science}\ }\textbf {\bibinfo {volume} {345}},\ \bibinfo {pages} {1135}
  (\bibinfo {year} {2014})}\BibitemShut {NoStop}%
\bibitem [{\citenamefont {DeCamp}\ \emph {et~al.}(2015)\citenamefont {DeCamp},
  \citenamefont {Redner}, \citenamefont {Baskaran}, \citenamefont {Hagan},\
  and\ \citenamefont {Dogic}}]{decamp2015orientational}%
  \BibitemOpen
  \bibfield  {author} {\bibinfo {author} {\bibfnamefont {S.~J.}\ \bibnamefont
  {DeCamp}}, \bibinfo {author} {\bibfnamefont {G.~S.}\ \bibnamefont {Redner}},
  \bibinfo {author} {\bibfnamefont {A.}~\bibnamefont {Baskaran}}, \bibinfo
  {author} {\bibfnamefont {M.~F.}\ \bibnamefont {Hagan}}, \ and\ \bibinfo
  {author} {\bibfnamefont {Z.}~\bibnamefont {Dogic}},\ }\href@noop {}
  {\bibfield  {journal} {\bibinfo  {journal} {Nat. Mater.}\ }\textbf {\bibinfo
  {volume} {14}},\ \bibinfo {pages} {1110} (\bibinfo {year}
  {2015})}\BibitemShut {NoStop}%
\bibitem [{\citenamefont {Giomi}\ \emph {et~al.}(2013)\citenamefont {Giomi},
  \citenamefont {Bowick}, \citenamefont {Ma},\ and\ \citenamefont
  {Marchetti}}]{giomi2013defect}%
  \BibitemOpen
  \bibfield  {author} {\bibinfo {author} {\bibfnamefont {L.}~\bibnamefont
  {Giomi}}, \bibinfo {author} {\bibfnamefont {M.~J.}\ \bibnamefont {Bowick}},
  \bibinfo {author} {\bibfnamefont {X.}~\bibnamefont {Ma}}, \ and\ \bibinfo
  {author} {\bibfnamefont {M.~C.}\ \bibnamefont {Marchetti}},\ }\href@noop {}
  {\bibfield  {journal} {\bibinfo  {journal} {Phys. Rev. Lett.}\ }\textbf
  {\bibinfo {volume} {110}},\ \bibinfo {pages} {228101} (\bibinfo {year}
  {2013})}\BibitemShut {NoStop}%
\bibitem [{\citenamefont {Thampi}\ \emph {et~al.}(2013)\citenamefont {Thampi},
  \citenamefont {Golestanian},\ and\ \citenamefont
  {Yeomans}}]{thampi2013velocity}%
  \BibitemOpen
  \bibfield  {author} {\bibinfo {author} {\bibfnamefont {S.~P.}\ \bibnamefont
  {Thampi}}, \bibinfo {author} {\bibfnamefont {R.}~\bibnamefont {Golestanian}},
  \ and\ \bibinfo {author} {\bibfnamefont {J.~M.}\ \bibnamefont {Yeomans}},\
  }\href@noop {} {\bibfield  {journal} {\bibinfo  {journal} {Phys. Rev. Lett.}\
  }\textbf {\bibinfo {volume} {111}},\ \bibinfo {pages} {118101} (\bibinfo
  {year} {2013})}\BibitemShut {NoStop}%
\bibitem [{\citenamefont {Giomi}\ \emph {et~al.}(2014)\citenamefont {Giomi},
  \citenamefont {Bowick}, \citenamefont {Mishra}, \citenamefont {Sknepnek},\
  and\ \citenamefont {Marchetti}}]{giomi2014defect}%
  \BibitemOpen
  \bibfield  {author} {\bibinfo {author} {\bibfnamefont {L.}~\bibnamefont
  {Giomi}}, \bibinfo {author} {\bibfnamefont {M.~J.}\ \bibnamefont {Bowick}},
  \bibinfo {author} {\bibfnamefont {P.}~\bibnamefont {Mishra}}, \bibinfo
  {author} {\bibfnamefont {R.}~\bibnamefont {Sknepnek}}, \ and\ \bibinfo
  {author} {\bibfnamefont {M.~C.}\ \bibnamefont {Marchetti}},\ }\href@noop {}
  {\bibfield  {journal} {\bibinfo  {journal} {Philos. Trans. Royal Soc. A}\
  }\textbf {\bibinfo {volume} {372}},\ \bibinfo {pages} {20130365} (\bibinfo
  {year} {2014})}\BibitemShut {NoStop}%
\bibitem [{\citenamefont {Thampi}\ \emph {et~al.}(2014)\citenamefont {Thampi},
  \citenamefont {Golestanian},\ and\ \citenamefont
  {Yeomans}}]{thampi2014instabilities}%
  \BibitemOpen
  \bibfield  {author} {\bibinfo {author} {\bibfnamefont {S.~P.}\ \bibnamefont
  {Thampi}}, \bibinfo {author} {\bibfnamefont {R.}~\bibnamefont {Golestanian}},
  \ and\ \bibinfo {author} {\bibfnamefont {J.~M.}\ \bibnamefont {Yeomans}},\
  }\href@noop {} {\bibfield  {journal} {\bibinfo  {journal} {Europhys. Lett.}\
  }\textbf {\bibinfo {volume} {105}},\ \bibinfo {pages} {18001} (\bibinfo
  {year} {2014})}\BibitemShut {NoStop}%
\bibitem [{\citenamefont {Alaimo}\ \emph {et~al.}(2017)\citenamefont {Alaimo},
  \citenamefont {K{\" o}hler},\ and\ \citenamefont
  {Voigt}}]{alaimo2017curvature}%
  \BibitemOpen
  \bibfield  {author} {\bibinfo {author} {\bibfnamefont {F.}~\bibnamefont
  {Alaimo}}, \bibinfo {author} {\bibfnamefont {C.}~\bibnamefont {K{\" o}hler}},
  \ and\ \bibinfo {author} {\bibfnamefont {A.}~\bibnamefont {Voigt}},\
  }\href@noop {} {\bibfield  {journal} {\bibinfo  {journal} {Sci. Rep.}\ }\textbf {\bibinfo {volume} {7}},\ \bibinfo {pages} {5211}
  (\bibinfo {year} {2017})}\BibitemShut {NoStop}%
\bibitem [{\citenamefont {Henkes}\ \emph {et~al.}(2017)\citenamefont {Henkes},
  \citenamefont {Marchetti},\ and\ \citenamefont
  {Sknepnek}}]{henkes2017dynamical}%
  \BibitemOpen
  \bibfield  {author} {\bibinfo {author} {\bibfnamefont {S.}~\bibnamefont
  {Henkes}}, \bibinfo {author} {\bibfnamefont {M.~C.}\ \bibnamefont
  {Marchetti}}, \ and\ \bibinfo {author} {\bibfnamefont {R.}~\bibnamefont
  {Sknepnek}},\ }\href@noop {} {\bibfield  {journal} {\bibinfo  {journal}
  {arXiv}\ }\textbf {\bibinfo {volume} {1705.05166}} (\bibinfo {year}
  {2017})}\BibitemShut {NoStop}%
\bibitem [{\citenamefont {Kaiser}(2003)}]{kaiser2003coupling}%
  \BibitemOpen
  \bibfield  {author} {\bibinfo {author} {\bibfnamefont {D.}~\bibnamefont
  {Kaiser}},\ }\href@noop {} {\bibfield  {journal} {\bibinfo  {journal} {Nat.
  Rev. Microbiol.}\ }\textbf {\bibinfo {volume} {1}},\ \bibinfo {pages}
  {45} (\bibinfo {year} {2003})}\BibitemShut {NoStop}%
\bibitem [{\citenamefont {Peruani}\ \emph {et~al.}(2012)\citenamefont
  {Peruani}, \citenamefont {Starru{\ss}}, \citenamefont {Jakovljevic},
  \citenamefont {S{\o}gaard-Andersen}, \citenamefont {Deutsch},\ and\
  \citenamefont {B{\"a}r}}]{peruani2012collective}%
  \BibitemOpen
  \bibfield  {author} {\bibinfo {author} {\bibfnamefont {F.}~\bibnamefont
  {Peruani}}, \bibinfo {author} {\bibfnamefont {J.}~\bibnamefont
  {Starru{\ss}}}, \bibinfo {author} {\bibfnamefont {V.}~\bibnamefont
  {Jakovljevic}}, \bibinfo {author} {\bibfnamefont {L.}~\bibnamefont
  {S{\o}gaard-Andersen}}, \bibinfo {author} {\bibfnamefont {A.}~\bibnamefont
  {Deutsch}}, \ and\ \bibinfo {author} {\bibfnamefont {M.}~\bibnamefont
  {B{\"a}r}},\ }\href@noop {} {\bibfield  {journal} {\bibinfo  {journal} {Phys.
  Rev. Lett.}\ }\textbf {\bibinfo {volume} {108}},\ \bibinfo {pages} {098102}
  (\bibinfo {year} {2012})}\BibitemShut {NoStop}%
\bibitem [{\citenamefont {Juelicher}\ \emph {et~al.}(2007)\citenamefont
  {Juelicher}, \citenamefont {Kruse}, \citenamefont {Prost},\ and\
  \citenamefont {Joanny}}]{juelicher2007active}%
  \BibitemOpen
  \bibfield  {author} {\bibinfo {author} {\bibfnamefont {F.}~\bibnamefont
  {Juelicher}}, \bibinfo {author} {\bibfnamefont {K.}~\bibnamefont {Kruse}},
  \bibinfo {author} {\bibfnamefont {J.}~\bibnamefont {Prost}}, \ and\ \bibinfo
  {author} {\bibfnamefont {J.-F.}\ \bibnamefont {Joanny}},\ }\href@noop {}
  {\bibfield  {journal} {\bibinfo  {journal} {Phys. Rep.}\ }\textbf
  {\bibinfo {volume} {449}},\ \bibinfo {pages} {3} (\bibinfo {year}
  {2007})}\BibitemShut {NoStop}%
\bibitem [{\citenamefont {Joanny}\ and\ \citenamefont
  {Prost}(2009)}]{joanny2009active}%
  \BibitemOpen
  \bibfield  {author} {\bibinfo {author} {\bibfnamefont {J.-F.}\ \bibnamefont
  {Joanny}}\ and\ \bibinfo {author} {\bibfnamefont {J.}~\bibnamefont {Prost}},\
  }\href@noop {} {\bibfield  {journal} {\bibinfo  {journal} {HFSP journal}\
  }\textbf {\bibinfo {volume} {3}},\ \bibinfo {pages} {94} (\bibinfo {year}
  {2009})}\BibitemShut {NoStop}%
\bibitem [{\citenamefont {Prost}\ \emph {et~al.}(2015)\citenamefont {Prost},
  \citenamefont {J{\"u}licher},\ and\ \citenamefont
  {Joanny}}]{prost2015active}%
  \BibitemOpen
  \bibfield  {author} {\bibinfo {author} {\bibfnamefont {J.}~\bibnamefont
  {Prost}}, \bibinfo {author} {\bibfnamefont {F.}~\bibnamefont {J{\"u}licher}},
  \ and\ \bibinfo {author} {\bibfnamefont {J.}~\bibnamefont {Joanny}},\
  }\href@noop {} {\bibfield  {journal} {\bibinfo  {journal} {Nat. Phys.}\
  }\textbf {\bibinfo {volume} {11}},\ \bibinfo {pages} {111} (\bibinfo {year}
  {2015})}\BibitemShut {NoStop}%
\bibitem [{\citenamefont {Gibbons}\ \emph {et~al.}(2001)\citenamefont
  {Gibbons}, \citenamefont {Chauwin}, \citenamefont {Desp{\'o}sito},\ and\
  \citenamefont {Jos{\'e}}}]{gibbons2001dynamical}%
  \BibitemOpen
  \bibfield  {author} {\bibinfo {author} {\bibfnamefont {F.}~\bibnamefont
  {Gibbons}}, \bibinfo {author} {\bibfnamefont {J.-F.}\ \bibnamefont
  {Chauwin}}, \bibinfo {author} {\bibfnamefont {M.}~\bibnamefont
  {Desp{\'o}sito}}, \ and\ \bibinfo {author} {\bibfnamefont {J.~V.}\
  \bibnamefont {Jos{\'e}}},\ }\href@noop {} {\bibfield  {journal} {\bibinfo
  {journal} {Biophys. J.}\ }\textbf {\bibinfo {volume} {80}},\ \bibinfo {pages}
  {2515} (\bibinfo {year} {2001})}\BibitemShut {NoStop}%
\bibitem [{\citenamefont {N{\'e}d{\'e}lec}(2002)}]{nedelec2002computer}%
  \BibitemOpen
  \bibfield  {author} {\bibinfo {author} {\bibfnamefont {F.}~\bibnamefont
  {N{\'e}d{\'e}lec}},\ }\href@noop {} {\bibfield  {journal} {\bibinfo
  {journal} {The Journal of cell biology}\ }\textbf {\bibinfo {volume} {158}},\
  \bibinfo {pages} {1005} (\bibinfo {year} {2002})}\BibitemShut {NoStop}%
\bibitem [{\citenamefont {Kraikivski}\ \emph {et~al.}(2006)\citenamefont
  {Kraikivski}, \citenamefont {Lipowsky},\ and\ \citenamefont
  {Kierfeld}}]{kraikivski2006enhanced}%
  \BibitemOpen
  \bibfield  {author} {\bibinfo {author} {\bibfnamefont {P.}~\bibnamefont
  {Kraikivski}}, \bibinfo {author} {\bibfnamefont {R.}~\bibnamefont
  {Lipowsky}}, \ and\ \bibinfo {author} {\bibfnamefont {J.}~\bibnamefont
  {Kierfeld}},\ }\href@noop {} {\bibfield  {journal} {\bibinfo  {journal}
  {Phys. Rev. Lett.}\ }\textbf {\bibinfo {volume} {96}},\ \bibinfo {pages}
  {258103} (\bibinfo {year} {2006})}\BibitemShut {NoStop}%
\bibitem [{\citenamefont {Kim}\ \emph {et~al.}(2009)\citenamefont {Kim},
  \citenamefont {Hwang}, \citenamefont {Lee},\ and\ \citenamefont
  {Kamm}}]{kim2009computational}%
  \BibitemOpen
  \bibfield  {author} {\bibinfo {author} {\bibfnamefont {T.}~\bibnamefont
  {Kim}}, \bibinfo {author} {\bibfnamefont {W.}~\bibnamefont {Hwang}}, \bibinfo
  {author} {\bibfnamefont {H.}~\bibnamefont {Lee}}, \ and\ \bibinfo {author}
  {\bibfnamefont {R.~D.}\ \bibnamefont {Kamm}},\ }\href@noop {} {\bibfield
  {journal} {\bibinfo  {journal} {PLoS Comput Biol}\ }\textbf {\bibinfo
  {volume} {5}},\ \bibinfo {pages} {e1000439} (\bibinfo {year}
  {2009})}\BibitemShut {NoStop}%
\bibitem [{\citenamefont {Jung}\ \emph {et~al.}(2015)\citenamefont {Jung},
  \citenamefont {Murrell},\ and\ \citenamefont {Kim}}]{jung2015f}%
  \BibitemOpen
  \bibfield  {author} {\bibinfo {author} {\bibfnamefont {W.}~\bibnamefont
  {Jung}}, \bibinfo {author} {\bibfnamefont {M.~P.}\ \bibnamefont {Murrell}}, \
  and\ \bibinfo {author} {\bibfnamefont {T.}~\bibnamefont {Kim}},\ }\href@noop
  {} {\bibfield  {journal} {\bibinfo  {journal} {Comput. Part.
  Mech.}\ }\textbf {\bibinfo {volume} {2}},\ \bibinfo {pages} {317}
  (\bibinfo {year} {2015})}\BibitemShut {NoStop}%
\bibitem [{\citenamefont {Popov}\ \emph {et~al.}(2016)\citenamefont {Popov},
  \citenamefont {Komianos},\ and\ \citenamefont {Papoian}}]{popov2016medyan}%
  \BibitemOpen
  \bibfield  {author} {\bibinfo {author} {\bibfnamefont {K.}~\bibnamefont
  {Popov}}, \bibinfo {author} {\bibfnamefont {J.}~\bibnamefont {Komianos}}, \
  and\ \bibinfo {author} {\bibfnamefont {G.~A.}\ \bibnamefont {Papoian}},\
  }\href@noop {} {\bibfield  {journal} {\bibinfo  {journal} {PLoS Comput Biol}\
  }\textbf {\bibinfo {volume} {12}},\ \bibinfo {pages} {e1004877} (\bibinfo
  {year} {2016})}\BibitemShut {NoStop}%
\bibitem [{\citenamefont {Jayaraman}\ \emph {et~al.}(2012)\citenamefont
  {Jayaraman}, \citenamefont {Ramachandran}, \citenamefont {Ghose},
  \citenamefont {Laskar}, \citenamefont {Bhamla}, \citenamefont {Kumar},\ and\
  \citenamefont {Adhikari}}]{gayathri-12}%
  \BibitemOpen
  \bibfield  {author} {\bibinfo {author} {\bibfnamefont {G.}~\bibnamefont
  {Jayaraman}}, \bibinfo {author} {\bibfnamefont {S.}~\bibnamefont
  {Ramachandran}}, \bibinfo {author} {\bibfnamefont {S.}~\bibnamefont {Ghose}},
  \bibinfo {author} {\bibfnamefont {A.}~\bibnamefont {Laskar}}, \bibinfo
  {author} {\bibfnamefont {M.~S.}\ \bibnamefont {Bhamla}}, \bibinfo {author}
  {\bibfnamefont {P.~B.~S.}\ \bibnamefont {Kumar}}, \ and\ \bibinfo {author}
  {\bibfnamefont {R.}~\bibnamefont {Adhikari}},\ }\href@noop {} {\bibfield
  {journal} {\bibinfo  {journal} {Phys. Rev. Lett.}\ }\textbf {\bibinfo
  {volume} {109}},\ \bibinfo {pages} {158302} (\bibinfo {year}
  {2012})}\BibitemShut {NoStop}%
\bibitem [{\citenamefont {Jiang}\ and\ \citenamefont {Hou}(2014)}]{hiang-14a}%
  \BibitemOpen
  \bibfield  {author} {\bibinfo {author} {\bibfnamefont {H.}~\bibnamefont
  {Jiang}}\ and\ \bibinfo {author} {\bibfnamefont {Z.}~\bibnamefont {Hou}},\
  }\href@noop {} {\bibfield  {journal} {\bibinfo  {journal} {Soft Matter}\
  }\textbf {\bibinfo {volume} {10}},\ \bibinfo {pages} {1012} (\bibinfo {year}
  {2014})}\BibitemShut {NoStop}%
\bibitem [{\citenamefont {Isele-Holder}\ \emph {et~al.}(2015)\citenamefont
  {Isele-Holder}, \citenamefont {Elgeti},\ and\ \citenamefont
  {Gompper}}]{riseleholder-15}%
  \BibitemOpen
  \bibfield  {author} {\bibinfo {author} {\bibfnamefont {R.~E.}\ \bibnamefont
  {Isele-Holder}}, \bibinfo {author} {\bibfnamefont {J.}~\bibnamefont
  {Elgeti}}, \ and\ \bibinfo {author} {\bibfnamefont {G.}~\bibnamefont
  {Gompper}},\ }\href@noop {} {\bibfield  {journal} {\bibinfo  {journal} {Soft
  Matter}\ }\textbf {\bibinfo {volume} {11}},\ \bibinfo {pages} {7181}
  (\bibinfo {year} {2015})}\BibitemShut {NoStop}%
\bibitem [{\citenamefont {Laskar}\ and\ \citenamefont
  {Adhikari}(2015)}]{laskar2015brownian}%
  \BibitemOpen
  \bibfield  {author} {\bibinfo {author} {\bibfnamefont {A.}~\bibnamefont
  {Laskar}}\ and\ \bibinfo {author} {\bibfnamefont {R.}~\bibnamefont
  {Adhikari}},\ }\href@noop {} {\bibfield  {journal} {\bibinfo  {journal} {Soft
  Matter}\ }\textbf {\bibinfo {volume} {11}},\ \bibinfo {pages} {9073}
  (\bibinfo {year} {2015})}\BibitemShut {NoStop}%
\bibitem [{\citenamefont {Kaiser}\ \emph {et~al.}(2015)\citenamefont {Kaiser},
  \citenamefont {Babel}, \citenamefont {ten Hagen}, \citenamefont {von
  Ferber},\ and\ \citenamefont {L{\"o}wen}}]{kaiser2015does}%
  \BibitemOpen
  \bibfield  {author} {\bibinfo {author} {\bibfnamefont {A.}~\bibnamefont
  {Kaiser}}, \bibinfo {author} {\bibfnamefont {S.}~\bibnamefont {Babel}},
  \bibinfo {author} {\bibfnamefont {B.}~\bibnamefont {ten Hagen}}, \bibinfo
  {author} {\bibfnamefont {C.}~\bibnamefont {von Ferber}}, \ and\ \bibinfo
  {author} {\bibfnamefont {H.}~\bibnamefont {L{\"o}wen}},\ }\href@noop {}
  {\bibfield  {journal} {\bibinfo  {journal} {J. Chem. Phys.}\
  }\textbf {\bibinfo {volume} {142}},\ \bibinfo {pages} {124905} (\bibinfo
  {year} {2015})}\BibitemShut {NoStop}%
\bibitem [{\citenamefont {Baskaran}\ and\ \citenamefont
  {Marchetti}(2008)}]{baskaran2008enhanced}%
  \BibitemOpen
  \bibfield  {author} {\bibinfo {author} {\bibfnamefont {A.}~\bibnamefont
  {Baskaran}}\ and\ \bibinfo {author} {\bibfnamefont {M.~C.}\ \bibnamefont
  {Marchetti}},\ }\href@noop {} {\bibfield  {journal} {\bibinfo  {journal}
  {Phys. Rev. Lett.}\ }\textbf {\bibinfo {volume} {101}},\ \bibinfo {pages}
  {268101} (\bibinfo {year} {2008})}\BibitemShut {NoStop}%
\bibitem [{\citenamefont {Peshkov}\ \emph {et~al.}(2012)\citenamefont
  {Peshkov}, \citenamefont {Aranson}, \citenamefont {Bertin}, \citenamefont
  {Chat\'e},\ and\ \citenamefont {Ginelli}}]{peshkov2012nonlinear}%
  \BibitemOpen
  \bibfield  {author} {\bibinfo {author} {\bibfnamefont {A.}~\bibnamefont
  {Peshkov}}, \bibinfo {author} {\bibfnamefont {I.~S.}\ \bibnamefont
  {Aranson}}, \bibinfo {author} {\bibfnamefont {E.}~\bibnamefont {Bertin}},
  \bibinfo {author} {\bibfnamefont {H.}~\bibnamefont {Chat\'e}}, \ and\
  \bibinfo {author} {\bibfnamefont {F.}~\bibnamefont {Ginelli}},\ }\href@noop
  {} {\bibfield  {journal} {\bibinfo  {journal} {Phys. Rev. Lett.}\ }\textbf
  {\bibinfo {volume} {109}},\ \bibinfo {pages} {268701} (\bibinfo {year}
  {2012})}\BibitemShut {NoStop}%
\bibitem [{\citenamefont {Gao}\ \emph {et~al.}(2015)\citenamefont {Gao},
  \citenamefont {Blackwell}, \citenamefont {Glaser}, \citenamefont
  {Betterton},\ and\ \citenamefont {Shelley}}]{gao2015multiscale}%
  \BibitemOpen
  \bibfield  {author} {\bibinfo {author} {\bibfnamefont {T.}~\bibnamefont
  {Gao}}, \bibinfo {author} {\bibfnamefont {R.}~\bibnamefont {Blackwell}},
  \bibinfo {author} {\bibfnamefont {M.~A.}\ \bibnamefont {Glaser}}, \bibinfo
  {author} {\bibfnamefont {M.~D.}\ \bibnamefont {Betterton}}, \ and\ \bibinfo
  {author} {\bibfnamefont {M.~J.}\ \bibnamefont {Shelley}},\ }\href@noop {}
  {\bibfield  {journal} {\bibinfo  {journal} {Phys. Rev. Lett.}\ }\textbf
  {\bibinfo {volume} {114}},\ \bibinfo {pages} {048101} (\bibinfo {year}
  {2015})}\BibitemShut {NoStop}%
\bibitem [{\citenamefont {Sekimoto}\ \emph {et~al.}(1995)\citenamefont
  {Sekimoto}, \citenamefont {Mori}, \citenamefont {Tawada},\ and\ \citenamefont
  {Toyoshima}}]{sekimoto-95}%
  \BibitemOpen
  \bibfield  {author} {\bibinfo {author} {\bibfnamefont {K.}~\bibnamefont
  {Sekimoto}}, \bibinfo {author} {\bibfnamefont {N.}~\bibnamefont {Mori}},
  \bibinfo {author} {\bibfnamefont {K.}~\bibnamefont {Tawada}}, \ and\ \bibinfo
  {author} {\bibfnamefont {Y.~Y.}\ \bibnamefont {Toyoshima}},\ }\href@noop {}
  {\bibfield  {journal} {\bibinfo  {journal} {Phys. Rev. Lett.}\ }\textbf
  {\bibinfo {volume} {75}},\ \bibinfo {pages} {172} (\bibinfo {year}
  {1995})}\BibitemShut {NoStop}%
\bibitem [{\citenamefont {Bourdieu}\ \emph {et~al.}(1995)\citenamefont
  {Bourdieu}, \citenamefont {Duke}, \citenamefont {Elowitz}, \citenamefont
  {Winkelmann}, \citenamefont {Leibler},\ and\ \citenamefont
  {Libchaber}}]{bourdieu-95}%
  \BibitemOpen
  \bibfield  {author} {\bibinfo {author} {\bibfnamefont {L.}~\bibnamefont
  {Bourdieu}}, \bibinfo {author} {\bibfnamefont {T.}~\bibnamefont {Duke}},
  \bibinfo {author} {\bibfnamefont {M.~B.}\ \bibnamefont {Elowitz}}, \bibinfo
  {author} {\bibfnamefont {D.~A.}\ \bibnamefont {Winkelmann}}, \bibinfo
  {author} {\bibfnamefont {S.}~\bibnamefont {Leibler}}, \ and\ \bibinfo
  {author} {\bibfnamefont {A.}~\bibnamefont {Libchaber}},\ }\href@noop {}
  {\bibfield  {journal} {\bibinfo  {journal} {Phys. Rev. Lett.}\ }\textbf
  {\bibinfo {volume} {75}},\ \bibinfo {pages} {176} (\bibinfo {year}
  {1995})}\BibitemShut {NoStop}%
\bibitem [{\citenamefont {Chelakkot}\ \emph {et~al.}(2013)\citenamefont
  {Chelakkot}, \citenamefont {Gopinath}, \citenamefont {Mahadevan},\ and\
  \citenamefont {Hagan}}]{rchelakkot-13}%
  \BibitemOpen
  \bibfield  {author} {\bibinfo {author} {\bibfnamefont {R.}~\bibnamefont
  {Chelakkot}}, \bibinfo {author} {\bibfnamefont {A.}~\bibnamefont {Gopinath}},
  \bibinfo {author} {\bibfnamefont {L.}~\bibnamefont {Mahadevan}}, \ and\
  \bibinfo {author} {\bibfnamefont {M.~F.}\ \bibnamefont {Hagan}},\ }\href@noop
  {} {\bibfield  {journal} {\bibinfo  {journal} {J. R. Soc. Interface}\
  }\textbf {\bibinfo {volume} {11}},\ \bibinfo {pages} {20130884} (\bibinfo
  {year} {2013})}\BibitemShut {NoStop}%
\bibitem [{\citenamefont {Liverpool}(2003)}]{tbliverpool-03}%
  \BibitemOpen
  \bibfield  {author} {\bibinfo {author} {\bibfnamefont {T.~B.}\ \bibnamefont
  {Liverpool}},\ }\href@noop {} {\bibfield  {journal} {\bibinfo  {journal}
  {Phys. Rev. E}\ }\textbf {\bibinfo {volume} {67}},\ \bibinfo {pages} {031909}
  (\bibinfo {year} {2003})}\BibitemShut {NoStop}%
\bibitem [{\citenamefont {Ghosh}\ and\ \citenamefont {Gov}(2014)}]{aghosh-14}%
  \BibitemOpen
  \bibfield  {author} {\bibinfo {author} {\bibfnamefont {A.}~\bibnamefont
  {Ghosh}}\ and\ \bibinfo {author} {\bibfnamefont {N.~S.}\ \bibnamefont
  {Gov}},\ }\href@noop {} {\bibfield  {journal} {\bibinfo  {journal} {Biophys.
  J.}\ }\textbf {\bibinfo {volume} {107}},\ \bibinfo {pages} {1065} (\bibinfo
  {year} {2014})}\BibitemShut {NoStop}%
\bibitem [{\citenamefont {Eisenstecken}\ \emph {et~al.}(2016)\citenamefont
  {Eisenstecken}, \citenamefont {Gompper},\ and\ \citenamefont
  {Winkler}}]{eisenstecken2016conformational}%
  \BibitemOpen
  \bibfield  {author} {\bibinfo {author} {\bibfnamefont {T.}~\bibnamefont
  {Eisenstecken}}, \bibinfo {author} {\bibfnamefont {G.}~\bibnamefont
  {Gompper}}, \ and\ \bibinfo {author} {\bibfnamefont {R.~G.}\ \bibnamefont
  {Winkler}},\ }\href@noop {} {\bibfield  {journal} {\bibinfo  {journal}
  {Polymers}\ }\textbf {\bibinfo {volume} {8}},\ \bibinfo {pages} {304}
  (\bibinfo {year} {2016})}\BibitemShut {NoStop}%
\bibitem [{\citenamefont {Cates}\ and\ \citenamefont
  {Tailleur}(2015)}]{cates2015motility}%
  \BibitemOpen
  \bibfield  {author} {\bibinfo {author} {\bibfnamefont {M.~E.}\ \bibnamefont
  {Cates}}\ and\ \bibinfo {author} {\bibfnamefont {J.}~\bibnamefont
  {Tailleur}},\ }\href@noop {} {\bibfield  {journal} {\bibinfo  {journal} {Ann.
  Rev. Cond. Matt. Phys.}\ }\textbf {\bibinfo {volume} {6}},\ \bibinfo {pages}
  {219} (\bibinfo {year} {2015})}\BibitemShut {NoStop}%
\bibitem [{\citenamefont {Redner}\ \emph {et~al.}(2013)\citenamefont {Redner},
  \citenamefont {Hagan},\ and\ \citenamefont {Baskaran}}]{redner2013structure}%
  \BibitemOpen
  \bibfield  {author} {\bibinfo {author} {\bibfnamefont {G.~S.}\ \bibnamefont
  {Redner}}, \bibinfo {author} {\bibfnamefont {M.~F.}\ \bibnamefont {Hagan}}, \
  and\ \bibinfo {author} {\bibfnamefont {A.}~\bibnamefont {Baskaran}},\
  }\href@noop {} {\bibfield  {journal} {\bibinfo  {journal} {Phys. Rev. Lett.}\
  }\textbf {\bibinfo {volume} {110}},\ \bibinfo {pages} {055701} (\bibinfo
  {year} {2013})}\BibitemShut {NoStop}%
\bibitem [{\citenamefont {Kremer}\ and\ \citenamefont
  {Grest}(1990)}]{kkremer-90}%
  \BibitemOpen
  \bibfield  {author} {\bibinfo {author} {\bibfnamefont {K.}~\bibnamefont
  {Kremer}}\ and\ \bibinfo {author} {\bibfnamefont {G.~S.}\ \bibnamefont
  {Grest}},\ }\href@noop {} {\bibfield  {journal} {\bibinfo  {journal} {J.
  Chem. Phys.}\ }\textbf {\bibinfo {volume} {92}},\ \bibinfo {pages} {5057}
  (\bibinfo {year} {1990})}\BibitemShut {NoStop}%
\bibitem [{\citenamefont {Rapaport}(2004)}]{rapaport2004art}%
  \BibitemOpen
  \bibfield  {author} {\bibinfo {author} {\bibfnamefont {D.~C.}\ \bibnamefont
  {Rapaport}},\ }\href@noop {} {\emph {\bibinfo {title} {The Art of Molecular
  Dynamics Simulation}}},\ \bibinfo {edition} {2nd}\ ed.\ (\bibinfo
  {publisher} {Cambridge University Press},\ \bibinfo {address} {New York, NY,
  USA},\ \bibinfo {year} {2004})\BibitemShut {NoStop}%
\bibitem [{\citenamefont {Weeks}\ \emph {et~al.}(1971)\citenamefont {Weeks},
  \citenamefont {Chandler},\ and\ \citenamefont {Andersen}}]{weeks1971role}%
  \BibitemOpen
  \bibfield  {author} {\bibinfo {author} {\bibfnamefont {J.~D.}\ \bibnamefont
  {Weeks}}, \bibinfo {author} {\bibfnamefont {D.}~\bibnamefont {Chandler}}, \
  and\ \bibinfo {author} {\bibfnamefont {H.~C.}\ \bibnamefont {Andersen}},\
  }\href@noop {} {\bibfield  {journal} {\bibinfo  {journal} {J. Chem. Phys.}\
  }\textbf {\bibinfo {volume} {54}},\ \bibinfo {pages} {5237} (\bibinfo {year}
  {1971})}\BibitemShut {NoStop}%
\bibitem [{\citenamefont {Plimpton}(1995)}]{splimpton-95}%
  \BibitemOpen
  \bibfield  {author} {\bibinfo {author} {\bibfnamefont {S.}~\bibnamefont
  {Plimpton}},\ }\href@noop {} {\bibfield  {journal} {\bibinfo  {journal} {J.
  Chem. Phys.}\ }\textbf {\bibinfo {volume} {117}},\ \bibinfo {pages} {1}
  (\bibinfo {year} {1995})}\BibitemShut {NoStop}%
\bibitem [{\citenamefont {Leimkuhler}\ and\ \citenamefont
  {Matthews}(2015)}]{leimkuhler2015molecular}%
  \BibitemOpen
  \bibfield  {author} {\bibinfo {author} {\bibfnamefont {B.}~\bibnamefont
  {Leimkuhler}}\ and\ \bibinfo {author} {\bibfnamefont {C.}~\bibnamefont
  {Matthews}},\ }\href@noop {} {\emph {\bibinfo {title} {Molecular Dynamics:
  With Deterministic and Stochastic Numerical Methods}}}\ (\bibinfo
  {publisher} {Springer},\ \bibinfo {year} {2015})\BibitemShut {NoStop}%
\bibitem [{sup()}]{supporting}%
  \BibitemOpen
  \href@noop {} {\enquote {\bibinfo {title} {Supplemental material available
  at: [url will be inserted by publisher]},}\ }\BibitemShut {NoStop}%
\bibitem [{\citenamefont {Zapotocky}\ \emph {et~al.}(1995)\citenamefont
  {Zapotocky}, \citenamefont {Goldbart},\ and\ \citenamefont
  {Goldenfeld}}]{zapotocky1995kinetics}%
  \BibitemOpen
  \bibfield  {author} {\bibinfo {author} {\bibfnamefont {M.}~\bibnamefont
  {Zapotocky}}, \bibinfo {author} {\bibfnamefont {P.~M.}\ \bibnamefont
  {Goldbart}}, \ and\ \bibinfo {author} {\bibfnamefont {N.}~\bibnamefont
  {Goldenfeld}},\ }\href@noop {} {\bibfield  {journal} {\bibinfo  {journal}
  {Phys. Rev. E}\ }\textbf {\bibinfo {volume} {51}},\ \bibinfo {pages} {1216}
  (\bibinfo {year} {1995})}\BibitemShut {NoStop}%
\bibitem [{\citenamefont {Shi}\ and\ \citenamefont
  {Ma}(2013)}]{shi2013topological}%
  \BibitemOpen
  \bibfield  {author} {\bibinfo {author} {\bibfnamefont {X.-q.}\ \bibnamefont
  {Shi}}\ and\ \bibinfo {author} {\bibfnamefont {Y.-q.}\ \bibnamefont {Ma}},\
  }\href@noop {} {\bibfield  {journal} {\bibinfo  {journal} {Nat.
  Commun.}\ }\textbf {\bibinfo {volume} {4}} (\bibinfo {year}
  {2013})}\BibitemShut {NoStop}%
\bibitem [{\citenamefont {Doi}\ and\ \citenamefont
  {Edwards}(1988)}]{doi1988theory}%
  \BibitemOpen
  \bibfield  {author} {\bibinfo {author} {\bibfnamefont {M.}~\bibnamefont
  {Doi}}\ and\ \bibinfo {author} {\bibfnamefont {S.~F.}\ \bibnamefont
  {Edwards}},\ }\href@noop {} {\emph {\bibinfo {title} {The theory of polymer
  dynamics}}},\ Vol.~\bibinfo {volume} {73}\ (\bibinfo  {publisher} {Oxford
  university press},\ \bibinfo {year} {1988})\BibitemShut {NoStop}%
\bibitem [{\citenamefont {Reichenbach}(1965)}]{reichenbach1965swarm}%
  \BibitemOpen
  \bibfield  {author} {\bibinfo {author} {\bibfnamefont {B.}~\bibnamefont
  {Reichenbach}},\ }\href@noop {} {\bibfield  {journal} {\bibinfo  {journal}
  {Ber. Dtsch. Bot. Ges.}\ }\textbf {\bibinfo {volume} {78}},\ \bibinfo {pages}
  {102} (\bibinfo {year} {1965})},\ \bibinfo {note} {{\it Myxobacteriales:
  Schwarmentfaltung und Bildung von Protocysten}, Institut fŸr den
  wissenschaftlichen Film, Gšttingen, 1966}\BibitemShut {NoStop}%
\bibitem [{\citenamefont {Drescher}\ \emph {et~al.}(2011)\citenamefont
  {Drescher}, \citenamefont {Dunkel}, \citenamefont {Cisneros}, \citenamefont
  {Ganguly},\ and\ \citenamefont {Goldstein}}]{drescher2011fluid}%
  \BibitemOpen
  \bibfield  {author} {\bibinfo {author} {\bibfnamefont {K.}~\bibnamefont
  {Drescher}}, \bibinfo {author} {\bibfnamefont {J.}~\bibnamefont {Dunkel}},
  \bibinfo {author} {\bibfnamefont {L.~H.}\ \bibnamefont {Cisneros}}, \bibinfo
  {author} {\bibfnamefont {S.}~\bibnamefont {Ganguly}}, \ and\ \bibinfo
  {author} {\bibfnamefont {R.~E.}\ \bibnamefont {Goldstein}},\ }\href@noop {}
  {\bibfield  {journal} {\bibinfo  {journal} {Pro. Nat. Acad. Sci. USA}\
  }\textbf {\bibinfo {volume} {108}},\ \bibinfo {pages} {10940} (\bibinfo
  {year} {2011})}\BibitemShut {NoStop}%
\bibitem [{\citenamefont {Shi}(2016)}]{lammps_bd}%
  \BibitemOpen
  \bibfield  {author} {\bibinfo {author} {\bibfnamefont {G.}~\bibnamefont
  {Shi}},\ }\href {https://github.com/anyuzx/Lammps_brownian/} {} (\bibinfo
  {year} {2016}),\ \bibinfo {note}
  {\url{https://github.com/anyuzx/Lammps_brownian/}}\BibitemShut {NoStop}%
\end{thebibliography}

%

\end{document}